%% file: main.tex
\documentclass[a4paper,11pt]{article}
\pdfoutput=1 
\usepackage{jcappub}
\usepackage{amsmath}
\usepackage{amssymb}
\usepackage{bm}
\usepackage{xcolor}
\usepackage{graphicx}
\usepackage{enumitem}
\usepackage{geometry}
\usepackage{xspace}
\usepackage{pdflscape}
\usepackage{placeins}
\usepackage{epstopdf}
\usepackage{comment}
\usepackage{booktabs}
\usepackage{array}
\makeatletter
\gdef\@fpheader{}
\g@addto@macro\bfseries{\boldmath}
\makeatother

\input{newcommands}

\newcommand{\Hc}[1]{\mathcal{H}}

\renewcommand{\Hc}{\mathcal{H}}

\def\doi{http://doi.org}

\date{today}

\title{Analysing Hubble Tension and Gravitational Waves for $f(Q,T)$ Gravity Theories }

\author[a]{Shreya Banerjee}
 
\author[a]{Aritrya Paul}

\affiliation[a]{Department of Physics,
Indian Institute of Technology (Indian School of Mines),
Dhanbad, Jharkhand-826004, India.}

\emailAdd{shreya@iitism.ac.in}
\emailAdd{22ms0028@iitism.ac.in}

\abstract{In this work, we examine viable models of $f(Q,T)$ gravity theories against observational data with the aim to constrain the parameter space of these models. We have analysed five different models of $f(Q,T)$ gravity and tested them against Type Ia supernovae, Cosmic Chronometer data, Baryon Acoustic Oscillations data and Pantheon data. We put stringent constraints on the $f(Q,T)$ gravity models, $f(Q,T) = Q^{n} +\beta T$ $(n=1,2,3)$, $f(Q,T)=-\alpha Q-\beta T^{2}$ and $f(Q,T)=Q^{-2}T^{2}$ along with other cosmological parameters such as deceleration parameter, equation of state parameter and demonstrate their alignment
with the $\Lambda CDM$ model and the observational data. We show that these models have the capability to alleviate the Hubble tension, by predicting the present value of the Hubble parameter close to $74$km/s/Mpc. $f(Q,T)$ gravity theory introduces alterations in the background evolution and imposes a
friction term in the propagation of gravitational waves, this phenomenon has also been examined. We have shown their agreement with the Gravitational Wave (GW) luminosity distance with the Electromagnetic (EM) counter part data from Advanced LIGO and Advanced VIRGO across different observing runs capturing coalescence of Binary Neutron Stars (BNS), mergers of Binary Black Holes (BBHs), and Neutron Star-Black Hole (NSBH) binaries with EM counterparts.}


\begin{document}

\maketitle
\vspace{0.6cm}

\section{Introduction}
From the Big Bang to the formation of elements, dark matter, and antimatter, fundamental questions about the universe are increasingly entering the scientific discourse. Many of these questions now have potential solutions. Concurrently, cosmological discoveries have revealed that the universe is expanding at an accelerated rate. This acceleration, supported by observations of distant Type Ia Supernovae (SNe Ia), is one of the most intriguing questions of recent years. Furthermore, these observations have shown that a mysterious component known as dark energy constitutes roughly 70\% of the universe.\\
General Relativity, as a geometrical theory, has provided a comprehensive explanation of observational evidence and has introduced new insights into the nature of space and time. However, despite its remarkable success, recent observations have raised questions about the validity of standard General Relativity, suggesting it may have limitations, especially at cosmic scales.\\
The Standard Cosmological model, that is also known as the $\Lambda CDM$ model (Lambda Cold Dark Matter model), is the prevailing theoretical skeleton that describes the vast-scale structure and evolution of this universe with great deal of accuracy. Understanding the evolution of the cosmos hinges upon our capacity to demonstrate how bodies move under the influence of gravity. Therefore, through understanding how spacetime bends, which can be determined by the distribution of matter, we can predict the future trajectory of particles. In line with the Standard Cosmological Model, the universe is predominantly composed of dark energy, dark matter, and ordinary matter. Dark energy is believed to drive the universe's accelerated expansion, while dark matter provides the gravitational framework for galaxy and large-scale structure formation. 
 The standard cosmological model has been hailed for its remarkable successes in demonstrating phenomena such as the existence of Cosmic Microwave Background Radiation (CMBR) and the observed accelerating expansion of the universe.\\
The cosmological constant is the most straightforward candidate for dark energy, yet it presents significant challenges, such as the coincidence problem and fine-tuning issues\cite{bull2016beyond,Weinberg:1988cp}. In the last few years, few other significant tensions with certain datasets such as Hubble and $\sigma_8$ have appeared. Hubble Tension refers to the discrepancy between the measurements of rate of expansion of the universe, known as Hubble constant. By utilizing the Planck satellite data to scrutinize the Cosmic Microwave Background
Radiation, which originates from the “very early” universe, a precise value for the expansion rate($H_0$) of $67.44 \pm 0.58$km/s/Mpc\cite{ade2016planck} has been derived. Again, by utilizing Cepheid variables and SNe Ia from the Hubble Space Telescope (HST) project, the present day value of expansion rate stood at $74.03 \pm 1.42$km/s/Mpc\cite{HST:2000azd}. Thus these measurements have yielded a different value of $H_0$, which is known as Hubble Tension. These challenges have prompted scientists to explore alternative theories of gravity that extend beyond the standard $\Lambda$CDM model. Rapid cosmic expansion can be explained through two primary approaches. The first involves modifying the energy-momentum tensor in Einstein's field equations, which includes theories involving scalar fields (like quintessence and phantom models)\cite{Caldwell:1999ew,Tsujikawa:2013fta,Wang:1999fa}, exotic equations of state (such as the Chaplygin gas)\cite{Bento:2002ps,Kamenshchik:2001cp}, and effects like bulk viscosity\cite{Brevik:2005bj,Padmanabhan:1987dg}. The second approach focuses on altering the geometry of spacetime within Einstein's equations, commonly referred to as Modified Theories of Gravity.\\
Modified gravity is a significant area of research in modern cosmology, aiming to provide a unified explanation for both the early and late epochs of the universe. One of the simplest theories within this framework is $f(R)$ gravity, where the gravitational dynamics are described by an arbitrary function of $R$, the Ricci scalar. Despite its simplicity, the additional degree of freedom introduced by the metric formulation in $f(R)$\cite{Buchdahl:1970ynr,Kamenshchik:2001cp,Starobinsky:2007hu} gravity can lead to conflicts with observational data. To address these challenges, alternative modified gravity theories have been proposed, including the $f(R,T)$\cite{Harko:2014pqa} theory, which incorporates a non-minimal interaction between matter and geometry, and the $f(R,G)$ \cite{elizalde2010lambdacdm}theory, which involves modifications based on the Gauss-Bonnet invariant. These theories aim to better align with observational data while still offering explanations for cosmic acceleration.\\
We know that GR is a successful theory of gravity to date, profoundly influencing both physics and cosmology. GR can be represented in three equivalent geometrical frameworks: the curvature representation, the teleparallel representation, and the symmetric teleparallel representation. 
In teleparallel theory, torsion is employed to describe gravitational phenomena, leading to what is known as the teleparallel equivalent of GR or $f(T)$ gravity. A novel approach, based on the third representation of GR, has given rise to a new theory known as symmetric teleparallel gravity, where the gravitational interaction is geometrically described using non-metricity $Q$. This quantity represents the variation in the length of a vector during parallel transport. Consequently, a new extended gravity theory called $f(Q)$ was formulated. The cosmological implications of $f(Q)$ gravity have been explored, with observational constraints considered\cite{jimenez2018coincident,harko2018coupling}. Studies show that while the cosmological evolution in $f(Q)$ gravity is similar to that in $\Lambda$CDM, deviations emerge at the perturbation level\cite{khyllep2021cosmological}.\\
Yixin et al.\cite{xu2020weyl} extended symmetric teleparallel gravity by introducing a coupling between the non-metricity scalar $Q$ and the trace of the energy-momentum tensor $T$, creating a modified gravity framework described by an arbitrary function $f(Q,T)$\cite{xu2019f}. This framework has been applied to various cosmological models, particularly those focusing on the universe's accelerated expansion\cite{najera2022cosmological,pradhan2021models,shiravand2022cosmological,singh2022cosmological,mandal2023cosmic}. Recent research indicates that $f(Q,T)$ gravity plays a crucial role in influencing tidal forces and equations of motion in the Newtonian limit\cite{yang2021geodesic}. Additionally, generalized metric theories like $f(Q,T)$ gravity can be directly compared with experimental results and Newtonian gravity, making the post-Newtonian limit an effective framework for aligning theoretical predictions with solar system observations. By comparing the predictions of Weyl-type $f(Q,T)$ gravity regarding tidal forces with astrophysical observations, we can gain deeper insights into gravitational interactions and their geometric interpretations.\\
In this work, we plan to study and examine $f(Q,T)$ gravity theories and constrain the parameter space using an extensive compilation of cosmological datasets through Monte Carlo Simulation. Through our analysis we aim to test our model in the context of Hubble tension and do additional checks in terms of the cosmological behaviour. 
The recent detection of gravitational waves(GWs) from the merger of compact binaries by LIGO-VIRGO \cite{krolak2021recent} have placed new challenges to modified gravity theories. We expect within the framework of modified gravity theories, the tensor perturbation equations will get modified, thereby affecting GW propagation. One way of quantifying this deviation is through the study of GW luminosity distance.
Gravitational wave luminosity distance $d_{L}^{gw}$ is the distance to an astronomical source measured based on the amplitude of the detected signals. It is derived from the observed strain amplitude of the GW
signal and the masses of the objects involved in the GW event. The distance measurement
is independent of the expansion of the universe and provides a direct measure of source’s
distance. The GW luminosity distance is crucial for studying the properties of GW sources and for cosmological studies. 
Modified gravity theories in the context of gravitational waves GW and GW luminosity distances have been actively studied in recent years. \cite{belgacem2018gravitational,fanizza2020comparing}. GWs has also been studied in the context of $f(T)$ gravity theory with the third-generation gravitational-wave detectors\cite{chen2024prospects}. In our present work, we also plan to test our $f(Q,T)$ models and the best fit parameter space in the context of gravitational wave LIGO-VIRGO and KAGRA data\cite{abbott2019gwtc,abbott2021gwtc,abbott2023gwtc,abbott2024gwtc}.

The plan of our work is the following. In Section \ref{sec2}, we briefly overview the $f(Q,T)$ gravity theory and discuss the modified background and gravitational wave equations. In Section \ref{sec3} we discuss the various datasets that we have worked with in this paper along with our data analysis technique. Then in Sections \ref{sec4} \& \ref{sec5} we have analysed different $f(Q,T)$ models against Hubble tension and GW. Finally in Section \ref{sec6} we have discussed the results and conclusions.

\section{$f(Q,T)$ gravity Theory} \label{sec2}
$f(Q,T)$ gravity extends the framework of symmetric teleparallel gravity. Here, the gravitational action $\mathcal{L}$ is determined by an arbitrary function $f$ which depends on both the Non-metricity scalar $Q$ and the Trace of the matter energy-momentum tensor $T$. The overall action for $f(Q, T)$ gravity is expressed as\cite{xu2019f}:
\begin{equation}\label{1}
S=\int d^{4}x\sqrt{-g}\left(\frac{1}{16\pi}f(Q,T)+\mathcal{L}_{m}\right)
\end{equation}
where  $\mathcal{L}_{m}$ for the matter Lagrangian and $g$ for the determinant of the metric. Moreover, the non-metricity scalar\cite{jimenez2018coincident} is defined as,
\begin{equation}\label{2}
Q=-g^{\alpha \beta}(L_{\delta \alpha}^{\gamma}L_{\beta \gamma}^{\delta}-L^{\gamma}_{\delta \gamma}L_{\alpha \beta}^{\delta})
\end{equation}
where $L_{\delta \alpha}^{\gamma}$ is  the deformation tensor\cite{hehl1976general} given by,
\begin{equation}\label{3}
L_{\delta \alpha}^{\gamma}=-\frac{1}{2}g^{\gamma \lambda}( \nabla_{\alpha} g_{\delta \lambda}+\nabla_{\delta} g_{\lambda \alpha}-\nabla_{\lambda}g_{\beta \delta })
\end{equation}
By varying the action in Eq.(\ref{1}) w.r.t the components of the metric tensor, we obtain the field equations of the $f(Q,T)$ gravity theory as,
\begin{multline}\label{4}
-\frac{2}{\sqrt{-g}}\nabla_{\gamma}(f_{Q}\sqrt{-g}P^{\gamma}_{\beta \alpha}) 
-\frac{1}{2}fg_{\alpha \beta} 
+ f_{T}(T_{\alpha \beta}+\Theta_{\alpha \beta})
\\
-f_{Q}(P_{\alpha \gamma \delta}Q^{\gamma \delta}_{\beta} 
- 2Q^{\gamma \delta}_{\alpha}P_{\gamma \delta \beta}) 
= 8\pi T_{\alpha \beta}
\end{multline}
where,
$Q_{\beta \alpha \gamma}=\nabla_{\beta}g_{\alpha \gamma}$ is the non-metricity tensor, $f_Q=\frac{\partial f}{\partial Q}$,$f_T=\frac{\partial f}{\partial T}$, $T_{\alpha \beta}=-\frac{2}{\sqrt{-g}}\frac{\delta (\sqrt{-g}L_m)}{\delta g^{\alpha \beta}}$\\
The stress-energy tensor, $\Theta_{\alpha \beta}=g^{\gamma \delta}\frac{\delta T_{\gamma \delta}}{\delta g^{\alpha \beta}}$ and $P^{\gamma}_{\alpha \beta}$ is the super potential\cite{xu2019f}, that is given by,
\begin{equation}\label{5}
P^{\gamma}_{\alpha \beta}=-\frac{1}{2}L^{\gamma}_{\alpha \beta}+\frac{1}{4}\left(Q^{\gamma}-\tilde{Q}^{\gamma}\right)g_{\alpha \beta}-\frac{1}{4}\delta^{\gamma}_{(\alpha}Q_{\beta)}
\end{equation}
where, $Q_{\gamma}\equiv Q_{\gamma \; \delta}^{\delta}$ and $\tilde{Q}^{\delta}_{\gamma \delta}$\\
In terms of super potential, the non-metricity scaler is given by, $Q=-Q_{\alpha \beta \gamma}P^{\alpha \beta \gamma}$

\subsection{Evolution of the Friedmann equations}
To derive the Friedmann equations, which describe the evolution of the universe, we start by assuming that the universe's matter content can be modeled as a perfect fluid. This perfect fluid is characterized by an energy-momentum tensor, which encapsulates the fluid's density, pressure, and the dynamics of its is given by, $T^{\alpha}_{\beta}=diag(-\rho, p,p,p)$. Also the non-metricity function $Q$ for such a metric is calculated and obtained as $Q=6H^{2}$. Then for the tensor $\Theta^{\alpha}_{\beta}$ we obtain the expression,
\begin{equation}\label{6}
\Theta^{\alpha}_{\beta}=\delta^{\alpha}_{\beta}p-2T^{\alpha}_{\beta}=diag(2\rho+p,-p,-p,-p)
\end{equation}
Now, the mathematical notations regarding this gravity model are
\begin{equation}\label{7}
F \equiv f_{Q}
\end{equation}
and,
\begin{equation}\label{8}
8\pi \tilde{G} \equiv f_{T}
\end{equation}
Where $f_{Q}$ denotes the derivative w.r.t. non-metricity 
 $Q$ and $f_{T}$ denotes the derivative w.r.t. Trace of energy-momentum tensor $T$.
Now, for studying the background dynamics, We consider the Friedmann-Lemaitre-Robertson-Walker(FLRW) metric, given as 
\begin{equation}\label{9}
ds^{2}=-dt^{2}+a^{2}(t)\left[dr^{2}+r^{2}(d\theta^{2}+sin^{2}\theta d\phi^{2})\right]
\end{equation}
where, $a(t)$ is the scale factor, which is a function of cosmic time.

From the field equation, we can easily find the Friedmann equations for $f(Q,T)$ gravity as\cite{xu2019f},
\begin{equation}\label{10}
8\pi \rho=\frac{f}{2}-6FH^{2}-\frac{2\tilde{G}}{1+\tilde{G}}(\dot{F}H+F\dot{H})
\end{equation}
\begin{equation}\label{11}
8\pi p=-\frac{f}{2}+6FH^{2}+2(\dot{F}H+F\dot{H})
\end{equation}
Combining the preceding two equations will give the evolutionary equation for the Hubble function $H(t)$, expressed as follows:
\begin{equation}\label{12}
\dot{H}+\frac{\dot{F}}{F}H=\frac{4\pi}{F}(1+\tilde{G})(\rho +p)
\end{equation}
To ensure broad applicability, we'll adopt a cosmological matter model governed by an equation of state represented as 
\begin{equation}\label{13}
p=\omega\rho 
\end{equation} 
where $\omega$ is equation of state parameter. This linear equation of state can describe the behavior of baryonic matter across varying densities, allowing it to describe conditions from the high-density scenarios akin to the early universe to the low-density conditions characteristic of the contemporary universe.\\
Combining Eq.\eqref{13}, Eq.(\ref{10}) and Eq.(\ref{12}) we obtain the general expression for density of matter as,
\begin{equation}\label{14}
\rho=\frac{f-12FH^{2}}{16\pi [1+(1+\omega)\tilde{G}]}
\end{equation}
To obtain the cosmological outcomes that directly align model predictions with observations, we introduce a independent variable the redshift $z$, instead of the time variable $t$, defined according to,
\begin{equation}\label{15}
1+z=\frac{1}{a}
\end{equation}
where we have assumed the present day value of scale factor as $a_{0}=1$. Therefore the following relation now holds true,
\begin{equation}\label{16}
\frac{d}{dt}=-(1+z)H(z)\frac{d}{dz}
\end{equation}
By using this relation, we will find the evolution of the cosmological parameters w.r.t. redshift. From Eq.(\ref{12}) the evolution of Hubble parameter w.r.t. redshift is given by,
\begin{equation}\label{17}
    \frac{dH}{dz}=\frac{\dot{F}H-4\pi(1+\omega)(1+\tilde{G})\rho}{(1+z)H(z)F}
\end{equation}

\subsection{GW propagation in $f(Q,T)$ gravity}
Let us first recall that, in GR, the free propagation of tensor perturbations in a FLRW background is described by\cite{najera2022cosmological},
\begin{equation}\label{18}
    h_{A}''+2\mathcal{H}h_{A}'+k^{2}h_{A}=0
\end{equation}
where $h_{A}(\eta ,k)$ are the Fourier modes of the Gravitational Wave(GW) amplitude, $A$ in the suffix denotes the two polarization states $+$ and $\times$ of the GWs, $\eta$ denotes the conformal time and $\mathcal{H}=a'/a$ represents the conformal Hubble parameter. We further introduce a field $\chi_{A}(\eta, k)$ through the following conformal transformation,
\begin{equation}\label{19}
    h_{A}(\eta, k)=\frac{1}{a(\eta)}\chi_{A}(\eta, k)
\end{equation}
So, Eq.(\ref{18}) becomes,
\begin{equation}\label{20}
    \chi''_{A}+(k^{2}-a''/a)\chi_{A}=0
\end{equation}
Both in matter dominance and in the recent Dark Energy dominated epoch $a''/a \approx 1/\eta^{2}$. For sub-Horizon modes $k\eta >> 1$, and therefore $a''/a$ can be neglected compared to $k^{2}$. 
\begin{equation}\label{21}
    \chi''_{A}+k^{2}\chi_{A}=0
\end{equation}
This indicates that the dispersion relation for tensor perturbations is $\omega = k$, meaning that GWs propagate at the speed of light. The factor $1/a$ in Eq.(\ref{19}) illustrates how the amplitude of GWs diminishes as they travel over cosmological distances from their source to the observer. For inspiraling binaries, this results in the standard relationship between the GW amplitude and the luminosity distance, expressed as $h_{A}(\eta, k) \propto 1/d_{L}(z)$. Altering the coefficient of the $k^{2}$ term in Eq.(\ref{18}) would change the propagation speed of GWs relative to the speed of light. However, the GW170817/GRB 170817A event has placed a stringent limit on such modifications, with $|c_{gw} - c|/c < \mathcal{O}(10^{-15})$ \cite{monitor2017gravitational}, effectively ruling out a significant portion of scalar-tensor and vector-tensor modifications of GR\cite{baker2019constraints,creminelli2017dark,ezquiaga2017dark,sakstein2017implications}.

Now, let's explore the propagation of GWs in $f(Q,T)$ gravity which is similar to one in $f(Q)$ gravity. This was expected since in vacuum the energy-momentum tensor is zero. This similarity arises because, in a vacuum, the energy-momentum tensor vanishes, leading to a zero trace of the energy-momentum tensor $(T=0)$. For a perturbed Einstein-Hilbert action and the energy-momentum tensor expanded to the second order in the metric, within the context of the FLRW background metric, the behavior of GWs in $f(Q,T)$ gravity,
\begin{equation*}
    g_{\mu\nu}=a^{2}(\eta)diag(1,-1,-1,-1)
\end{equation*}
and by considering the TT gauge, Eq.(\ref{18}), the propagation equation for GWs in $f(Q,T)$ gravity will be, 
\begin{equation}\label{22}
h_{A}''+2\mathcal{H}\left[1-\delta(\eta)\right]h_{A}'+k^{2}h_{A}=0
\end{equation}
where $\delta(\eta)$ parametrizes the deviation from GR (also known as the friction term), which reads as
\begin{equation}\label{23}
\delta(\eta)=-\frac{1}{2\mathcal{H}}\frac{d(log f_{Q})}{d\eta}
\end{equation}
Like before, we introduce $\chi_{A}(\eta, k)$ as
\begin{equation}\label{24}
    h_{A}(\eta,k)=\frac{1}{\tilde{a}(\eta)}\chi_{A}(\eta,k)
\end{equation}
and we get $\chi''_{A}+(k^{2}-\tilde{a}''/\tilde{a})\chi_{A}=0$. Once again, inside the horizon the term $\tilde{a}''/\tilde{a}$ is totally negligible, so GWs propagate at the speed of light. $\tilde{a}$ is a modified or rescaled version of the scale factor, which appears in the equation to account for the effects of the universe's expansion on the perturbation $\chi_{A}$. However, while propagating across cosmological distances, $h_{A}$ now decreases as $1/\tilde{a}$ rather than $1/a$. 
The friction term present in Eq.(\ref{22}) alters the evolution of GW amplitude as it propagates through cosmological distances. In GR, the amplitude of a coalescing binary is inversely proportional to the luminosity distance as, $1/d_{L}$. This relationship causes a bias in the inferred luminosity distance derived from GW observations. Specifically, if $\delta(\eta) < 0$, the damping term is stronger, causing the GW amplitude to decrease more significantly during propagation from the source to the detector. Within the framework of GR, this would lead to the erroneous interpretation that the source is farther away than it actually is. Conversely, if $\delta(\eta) > 0$, the GW amplitude would appear larger, suggesting a source distance that is closer than its true distance. Therefore, it is essential to distinguish between two types of luminosity distances: the ``electromagnetic luminosity distance'', denoted as $d_{L}^{em}$, and the ``GW luminosity distance'', denoted as $d_{L}^{gw}$.
Then in such a modified gravity model, for a coalescing binary at redshift $z$, the two are now related as
\begin{equation}\label{25}
    d_{L}^{gw}(z)=\frac{a(z)}{\tilde{a}(z)}d_{L}^{em}=\frac{1}{(1+z)\tilde{a}(z)}d_{L}^{em}
\end{equation}
By substituting the scale factor $\mathcal{H}[1-\delta(\eta)]=\frac{\tilde{a'}}{\tilde{a}}$ and performing some straightforward calculation, we can redefine the relation between GW luminosity distance and EM luminosity distance as,
\begin{equation}\label{26}
d_{L}^{gw}=d_{L}^{em}exp\left(-\int_{0}^{z}\frac{dz'}{(1+z')\delta(z')}\right)
\end{equation}
Thus we see that the difference between the two types of luminosity distances now depends on the friction term which directly depends on the corresponding modified gravity model. By quantifying deviations of the GW luminosity distance concerning the electromagnetic luminosity distance, we would be able to measure deviations from GR. 

In $f(Q,T)$ gravity model, the above relation takes the
form,
\begin{equation}\label{27}
    d_{L}^{gw}=\sqrt{\frac{f_{Q}(0)}{f_{Q}}}d_{L}^{em}
\end{equation}
Here $f_{Q}(0)$ is a function $f_{Q}$ computed at present time i.e. $z=0$. From the above equation and using EM wave luminosity distance 
\begin{equation}\label{28}
    d^{em}_{L}(z)=c(1+z)\int_{0}^{z}\frac{dz'}{H(z')}
\end{equation} 
we can obtain the expression of $d_{L}^{gw}$ in terms of redshift $z$. as,
\begin{equation}\label{29}
    d_{L}^{gw}=c(1+z)\sqrt{\frac{f_{Q}(0)}{f_{Q}}} \int_{0}^{z}\frac{dz'}{H(z')}
\end{equation}

\section{Current observational data} \label{sec3}
The $f(Q,T)$ gravity theory offers a compelling framework for explaining cosmic evolution across various facets, significantly impacting the structure formation in the early universe. By altering the background dynamics and modifying the gravitational wave luminosity distance, this theory presents a promising approach to addressing the Hubble Tension and analyzing gravitational wave data. In this study, we utilize a range of observational datasets, including Cosmic Chronometer data\cite{yu2018hubble}, Pantheon+ Type Ia supernovae (SN Ia) data\cite{brout2022pantheon+}, Union-2.1 SN Ia samples\cite{suzuki2012hubble}, BAO observations\cite{naik2023impact}, and gravitational wave data from LIGO and VIRGO\cite{abbott2019gwtc,abbott2021gwtc,abbott2023gwtc,abbott2024gwtc}, to constrain the parameters of $f(Q,T)$ gravity models. To evaluate these datasets, we employ Bayesian statistical analysis and utilize the $emcee$\cite{foreman2013emcee} package in Python to perform a Markov Chain Monte Carlo (MCMC) method.\\
First, we examine the priors on the parameters, as presented in the tables. For our MCMC analysis, we used 100 walkers with 500 or 1000 iterations across all datasets. The following subsection provides a more detailed discussion of the datasets and the statistical analysis performed.
\subsection{Cosmic Chronometer(CC) Dataset}
The cosmological principle, a fundamental assumption in cosmology, states that the universe is homogeneous and isotropic when viewed on a large scale. This principle plays a crucial role in observational cosmology, where the Hubble parameter $H = \frac{\dot{a}}{a}$ is utilized to directly examine the universe's expansion. Here, $a$ represents the scale factor, and $\dot{a}$ is its time derivative. This relationship forms the basis for analyzing the expansion of the universe within the context of the FLRW metric.

The CC method enables the measurement of the Hubble parameter $H(z)$ independently of any specific cosmological assumptions. CC data is based on the $H(z)$ measurement thorough the relative ages of passively evolving galaxies and the corresponding estimation of $dz/dt$\cite{jimenez2002constraining}. In this work we used 31 points $z-H(z)$ cosmic chronometer data from various sources\cite{jimenez2003constraints,simon2005constraints,stern2010cosmic, moresco2012improved,zhang2014four,xia2016constraining,moresco20166,ratsimbazafy2017age} using cosmic chronometer(CC) method in the redshift range $0.0708 < z < 1.965$.
Further, we used the below chi-square function to obtain the best-fit values of the model parameters (which is equal to the maximum likelihood function),
\begin{equation}\label{30}
\chi^{2}_{CC}=\Sigma_{n=1}^{31}\frac{[H_{i}^{th}(\Theta_{i}, z_{i})-H_{i}^{obs}(z_{i})]^{2}}{\sigma_{Hubble}^{2}(z_{i})}
\end{equation}
where $H_{i}^{th}$ is the theoretical value of the Hubble parameter with model parameters $\Theta_{i}$, $H_{i}^{obs}$ denotes the observed value and $\sigma_{Hubble}^{2}$ denotes the standard error in the observed value of $H_{CC}(z)$.
For the CC dataset, the chi-square function, $\chi^{2}$ is \\
\begin{equation}\label{31}
\chi^{2}=\chi^{2}_{CC}
\end{equation}

\subsection{Baryon Acoustic Oscillation(BAO) Dataset}

Baryonic Acoustic Oscillations (BAOs) are a key cosmological tool for investigating the large-scale structure of the Universe. These oscillations arise from acoustic waves in the early universe, which caused compression in the photon-baryon fluid. This compression created a characteristic peak in the galaxy correlation function, acting as a standard ruler for cosmic distance measurements. The comoving size of the BAO peak is determined by the sound horizon at recombination, which is influenced by the baryon density and the temperature of the cosmic microwave background.\\
At a given redshift $z$, the position of the BAO peak in the angular direction determines the angular separation $\Delta \theta=r_{d}/(1+z)D_{A}(z)$, while in the radial direction, it determines the redshift separation $\Delta z=r_{d}/D_{H}(z)$. Here, $D_{A}$ is the angular distance, $D_{H}=c/H$ corresponds to the Hubble distance, and $r_{d}$ represents the sound horizon at the drag epoch. By  determining the position of the BAO peak at different redshifts, we can constrain combinations of cosmological parameters that determine $D_{H}/r_{d}$ and $D_{A}/r_{d}$. By selecting an appropriate value for $r_{d}$, we can estimate $H(z)$\cite{naik2023impact}. In this work, we considered a dataset containing 26 data points obtained from line-of-sight BAO measurements\cite{delubac2013baryon,gaztanaga2009clustering,blake2012wigglez}.\\
Similar to the cosmic chronometers(CC) method, the chi-squared for BAO data is,
\begin{equation}\label{32}
    \chi^{2}_{BAO}=\Sigma_{n=1}^{26}\frac{[H_{i}^{th}(\Theta_{i}, z_{i})-H_{i}^{obs}(z_{i})]^{2}}{\sigma_{BAO}^{2}(z_{i})}
\end{equation}
where, like before, $H_{i}^{th}$ is the theoretical value of the Hubble parameter with model parameters $\Theta_{i}$, $H_{i}^{obs}$ denotes the observed value and $\sigma_{BAO}^{2}$ denotes the standard error in the observed value of $H_{BAO}(z)$.
For the CC dataset and BAO dataset, the total chi-square function is,
\begin{equation}\label{33}
\chi_{T}^{2}=\chi^{2}_{CC}+\chi^{2}_{BAO}
\end{equation}
The combined analysis allows for a more comprehensive constraint on the model parameters by incorporating the information from the cosmic chronometers(CC) and baryonic acoustic oscillations(BAO).

\subsection{Type Ia Supernova(SN Ia) Datasets}
Type Ia supernovae are widely recognized as standard candles\cite{leibundgut2000type,hillebrandt2000type} for measuring cosmic acceleration in the local universe. In our analysis we have used the Pantheon+ dataset\cite{brout2022pantheon+} and Union-2.1 dataset\cite{suzuki2012hubble}. The Pantheon+(SN Ia) dataset contains distance modulus data spanning within a redshift range of $0.00122 < z < 2.26137$, acquired from 18 distinct surveys. The Union 2.1 dataset contains 580 data points of SN Ia with a redshift range of $0.015 < z < 1.414$.\\
In this paper, we compare the theoretical value $\mu^{th}_{i}$ with the measured value $\mu^{obs}_{i}$ of the distance modulus to estimate our model parameters.\\
The theoretical distance modulus is defined as follows:
\begin{equation}\label{34}
\mu=m-M=5\;log_{10}[d_{L}(z)]+25  
\end{equation}
where, $m$ and $M$ denotes the apparent and absolute magnitude and $d_{L}(z)$ is the luminosity distance which can be determined using the following formula,
\begin{equation}\label{35}
d_{L}(z)=c(1+z)\int_{0}^{z}\frac{dz'}{H(z')}
\end{equation}
where $H(z)$ is Hubble function described by the corresponding models and $c$ is the speed of light.
The chi-square function for the Pantheon dataset is defined as,
\begin{equation}\label{36}
\chi^{2}_{Panth}=\Sigma_{i,j=1}^{1701}\Delta \mu_{i}(C_{Panth}^{-1})\Delta \mu_{j}
\end{equation}
Here $C_{Panth}$ is covariance matrix, and $\Delta \mu_{i}=\mu^{th}(\Theta_{i}, z_{i})-\mu^{obs}(z_{i})$ is the difference between the theoretical value determined from the model with model parameters $\Theta_{i}$ and observed distance modulus value obtained from cosmic data and. Similarly the $\chi^{2}$ function for Union 2.1 dataset is given by,
\begin{equation}\label{37}
    \chi^{2}_{Union\;2.1}=\Sigma_{n=1}^{580}\frac{[\mu_{i}^{th}(\Theta_{i}, z_{i})-\mu_{i}^{obs}(z_{i})]^{2}}{\sigma_{Union\;2.1}^{2}(z_{i})}
\end{equation}
where $\mu_{i}^{th}$ is the theoretical value of the distance modulus with model parameters $\Theta_{i}$, $\mu_{i}^{obs}$ denotes the observed value and $\sigma_{Union\;2.1}^{2}$ denotes the standard error in the observed value of $\mu_{union}(z)$.

Now the total chi-square function is as follows:
\begin{equation}\label{38}
\chi^{2}_{T}=\chi^{2}_{CC}+\chi^{2}_{BAO}+\chi^{2}_{Panth}+\chi^{2}_{Union\;2.1}   
\end{equation}

\subsection{Gravitational Wave(GW) Dataset}
As a new window into the Universe, GW signals offer unique opportunities. Specifically, GWs emitted by inspiraling binary systems—such as Binary Black Holes (BH-BH), Neutron Stars (NS-NS), or mixed Neutron Star-Black Hole (NS-BH) pairs—can serve as `standard sirens', providing direct measurements of luminosity distances without relying on the cosmic distance ladder\cite{desmond2019local}. Unlike observations of SN Ia in the electromagnetic (EM) domain, the major advantage of GWs lies in their independent calibration of luminosity distances. Recent studies have explored the potential of extending cosmic curvature tests using simulated GW data from next-generation GW detectors, including third-generation ground-based detectors like the Einstein Telescope (ET)\cite{sathyaprakash2012scientific} and Cosmic Explorer (CE)\cite{hall2022cosmic}, as well as space-based detectors like LISA\cite{amaro2017laser}.
In this work we have used LIGO-VIRGO and KAGRA collaboration(Gravitational-Wave-Transient-Catalog) GWTC-1\cite{abbott2019gwtc}, GWTC-2\cite{abbott2021gwtc}, GWTC-2.1\cite{abbott2024gwtc} and GWTC-3\cite{abbott2023gwtc} of compact binary coalescences observed by LIGO and VIRGO during their different observing run. In our analysis we have used 138 GW luminosity distance($d_{L}^{gw}$) vs. redshift $z$ data points spanning with in a redshift range $0.01 < z < 1.18$.

\section{Hubble Tension} \label{sec4}
In this present section we will probe some specific cosmological models in $f(Q,T)$ gravity theory and for sake of generality we will assume that the cosmological matter follows the following equation of state (EOS) $p=\omega\rho$,
where $\omega$ is equation of state parameter.\\
Now, we will report the results that we have obtained from the observational data analysis of the models with the different datasets. As mentioned earlier, we have used $emcce$\cite{foreman2013emcee} Python package and MCMC technique for visualization and the subsequent analysis. Using MCMC, we calculated the bounds of the free parameters in 68\%, 95\% credible level. The probability function used to maximize the best fits of the parameters is,\\
\begin{equation}\label{39}
\mathcal{L} \propto exp(-\chi^{2}/2)
\end{equation}
where $\chi^{2}$ is the chi-squared function.

We examined our models by assuming a specific parameterization form of EoS as a function of redshift $z$,
\begin{equation}\label{40}
    \omega (z)=-\frac{1}{1+m(1+z)^{3}}
\end{equation}
\noindent
Where $m$ is a free parameter. The assumed form of $\omega(z)$ is chosen in way so that at very large redshift ($z \gg 1$), corresponding to the early phase of the universe, $\omega$ is approximately zero. This reflects the Equation of State (EoS) parameter for a pressureless fluid, such as ordinary matter. However, as the universe evolves and the redshift decreases to the present time ($z = 0$), $\omega$ gradually shifts to a negative value. At $z = 0$, this results in a negative value of $\omega = -\frac{1}{1+m}$. 

    In the following subsections, we discuss the $f(Q,T)$ models with which we have performed our analysis.

\subsection{Model : $f(Q,T)= Q^{n}+\beta T$}
\label{model I}
As a example, we will consider the following minimally coupled cosmological model having non-linear dependence on $Q$. We consider $f(Q,T)= Q^{n}+\beta T$, where $\beta$ is a constant. For this case we have,
\begin{equation}\label{40.1} 
F=f_{Q}=nQ^{n-1}=n6^{n-1}H^{2n-2}, \;\;\; 8\pi \tilde{G}=f_{T}=\beta
\end{equation}
Now, solving for $p$ and $\rho$ from Friedmann equations (\ref{10}) and (\ref{11}) for this particular form, we obtain the equation of state parameter $\omega =\frac{p}{\rho}$ as,
\begin{equation}\label{40.2}
    \omega=\frac{(16\pi+3\beta)\dot{H}+\frac{3}{n}(8\pi+\beta)H^{2}}{\beta \dot{H}-\frac{3}{n}(8\pi+\beta)H^{2}}
\end{equation}
By using Eq.(\ref{16}) and Eq.(\ref{40}) we obtain the following differential equation for Hubble parameter, 
\begin{equation}\label{40.3}
    \frac{dH}{dz}=\frac{3m(8\pi+\beta)(1+z)^{2}H(z)}{n[\beta+(16\pi+3\beta)(1+m(1+z)^{3})]}
\end{equation}
Solving the above equation we obtain the evolution of the Hubble parameter w.r.t. $z$ as,
\begin{equation}\label{40.4}
    H(z)=H_{0}\left(\frac{\beta + (16\pi+3\beta)(1+m(1+z)^{3})}{\beta+(16\pi+3\beta)(1+m)}\right)^{l}
\end{equation}
where $l=\frac{(8\pi+\beta)}{n(16\pi+3\beta)}$ and $H_{0}$ is Hubble parameter value at present day, i.e at $z=0$.

We now analyse the following three sub cases for this model corresponding to $n=1,\ 2,\ 3$.

\subsubsection{Model I : $f(Q,T)= Q+\beta T$}
As first example of our cosmological model in $f(Q,T)$ gravity we will consider minimally coupled model with linear dependence on $Q$, $T$. We consider the following form
\begin{equation}\label{41}
  f(Q,T)= Q+\beta T
\end{equation}
where $\beta$ is a constant. 

\begin{figure}[h!]
\centering
\includegraphics[scale=0.6]{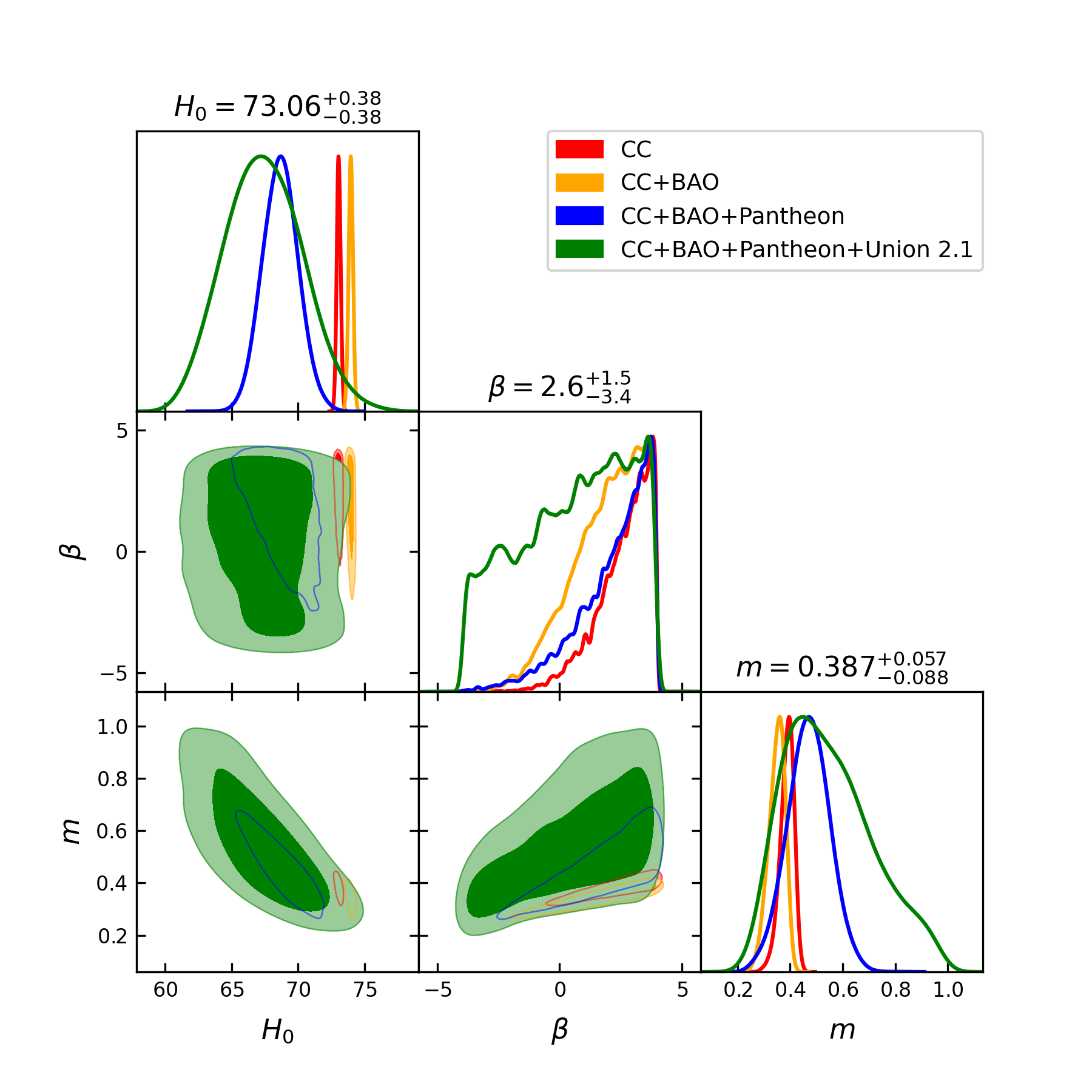}
\caption{\label{n=0_(4_legended_plot)}2D-contour plot of the model parameters ($H_0$,$\beta$,$m$) of Model I. The deeper shade show 68\% credible level (C.L.) and the lighter shade show 95\% credible level (C.L.)} 
\end{figure}

Then we immediately obtain,
\begin{equation}\label{42}
F=f_{Q}=1 \; ; \;8\pi \tilde{G}=f_{T}=\beta
\end{equation}
Now, solving for $p$ and $\rho$ from Friedmann equations (\ref{10}) and (\ref{11}) for this particular form, we obtain the equation of state parameter $\omega =\frac{p}{\rho}$ as,
\begin{equation}\label{43}
    \omega=\frac{(16\pi+3\beta)\dot{H}+3(8\pi+\beta)H^{2}}{\beta \dot{H}-3(8\pi+\beta)H^{2}}
\end{equation}
By using Eq.(\ref{16}) and Eq.(\ref{40}) we obtain the following differential equation for Hubble parameter, 
\begin{equation}\label{44}
    \frac{dH}{dz}=\frac{3m(8\pi+\beta)(1+z)^{2}H(z)}{[\beta+(16\pi+3\beta)(1+m(1+z)^{3})]}
\end{equation}
Solving the above equation we obtain the evolution of the Hubble parameter w.r.t. $z$ as,
\begin{equation}\label{45}
    H(z)=H_{0}\left(\frac{\beta + (16\pi+3\beta)(1+m(1+z)^{3})}{\beta+(16\pi+3\beta)(1+m)}\right)^{l}
\end{equation}
where $l=\frac{(8\pi+\beta)}{(16\pi+3\beta)}$ and $H_{0}$ is Hubble parameter value at present day, i.e at $z=0$.\\
Using Eq. \eqref{45}, we perform the data analysis for the datasets as explained in Section \ref{sec3}. The parameters that we want to constrain are $(H_0,\ \beta,\ m)$.
Fig \ref{n=0_(4_legended_plot)} contour plot show the the best fit values for the parameters of this model. The best fit values of this model parameters are shown in Table \ref{Table_1} for different datasets and the joint analysis along with the priors. Interestingly we see that the joint analysis of CC+BAO+Pantheon and CC+BAO+Pantheon+Union gives the present value of the Hubble parameter as $73.97_{-0.45}^{+0.44}$ and $73.06_{-0.38}^{+0.38}$ respectively, as par with the direct measurement of the 2019 SH0ES collaboration ($H_0 = (74.03 \pm 1.42)$ km/s/Mpc) \cite{Riess:2019cxk}. Thus this model clearly alleviates the existing tension related to the present value of the Hubble parameter without the need for any cosmological constant. The corresponding best fit values of ($\beta,\ m$) are ($1.9_{-4.3}^{+2.1},\ 0.348_{-0.097}^{+0.065} $) and ($2.6_{-3.4}^{+1.5},\ 0.387_{-0.088}^{+0.057} $) respectively. In Fig.\ref{distance modulus(Pantheon_data,n=0)} we have plotted our model and $\Lambda CDM$ model with the best fit model parameters along with the Pantheon+ SNe Ia data, with their error bars. As we can see, our model fits the data as good as $\Lambda$CDM. \\
\begin{table*}[h!]
\caption{\label{Table_1}Best fit parameter values for Model I from MCMC}
\vspace{2mm}
\begin{tabular}{m{5cm} m{3.5cm} m{2.5cm} m{2cm}}
\hline
 Datasets & $H_{0}(km/s/Mpc)$ & $\beta$ & $m$ \\ 
\hline 
\vspace{1mm}
Priors & (60,80) & (-4,4) & (0,1)\\ 
\vspace{1mm}
 CC & $67_{-7}^{+8}$ & $0.6_{-4.4}^{+3.5}$ & $0.55_{-0.32}^{+0.44}$ \\
 \vspace{1mm}
 CC+BAO & $68_{-3.4}^{+3.6}$ & $2.3_{-4.7}^{+1.8}$ & $0.47_{-0.22}^{+0.21}$ \\
 \vspace{1mm}
 CC+BAO+Pantheon & $73.97_{-0.45}^{+0.44}$ & $1.9_{-4.3}^{+2.1}$ & $0.348_{-0.097}^{+0.065}$ \\
 \vspace{1mm}
CC+BAO+Pantheon+Union & $73.06_{-0.38}^{+0.38}$ & $2.6_{-3.4}^{+1.5}$ & $0.387_{-0.088}^{+0.057}$ \\
\hline
\end{tabular}
\end{table*}

\begin{figure}[htbp]
    \centering
    \begin{minipage}[b]{0.46\textwidth}
        \centering
        \includegraphics[width=\textwidth]{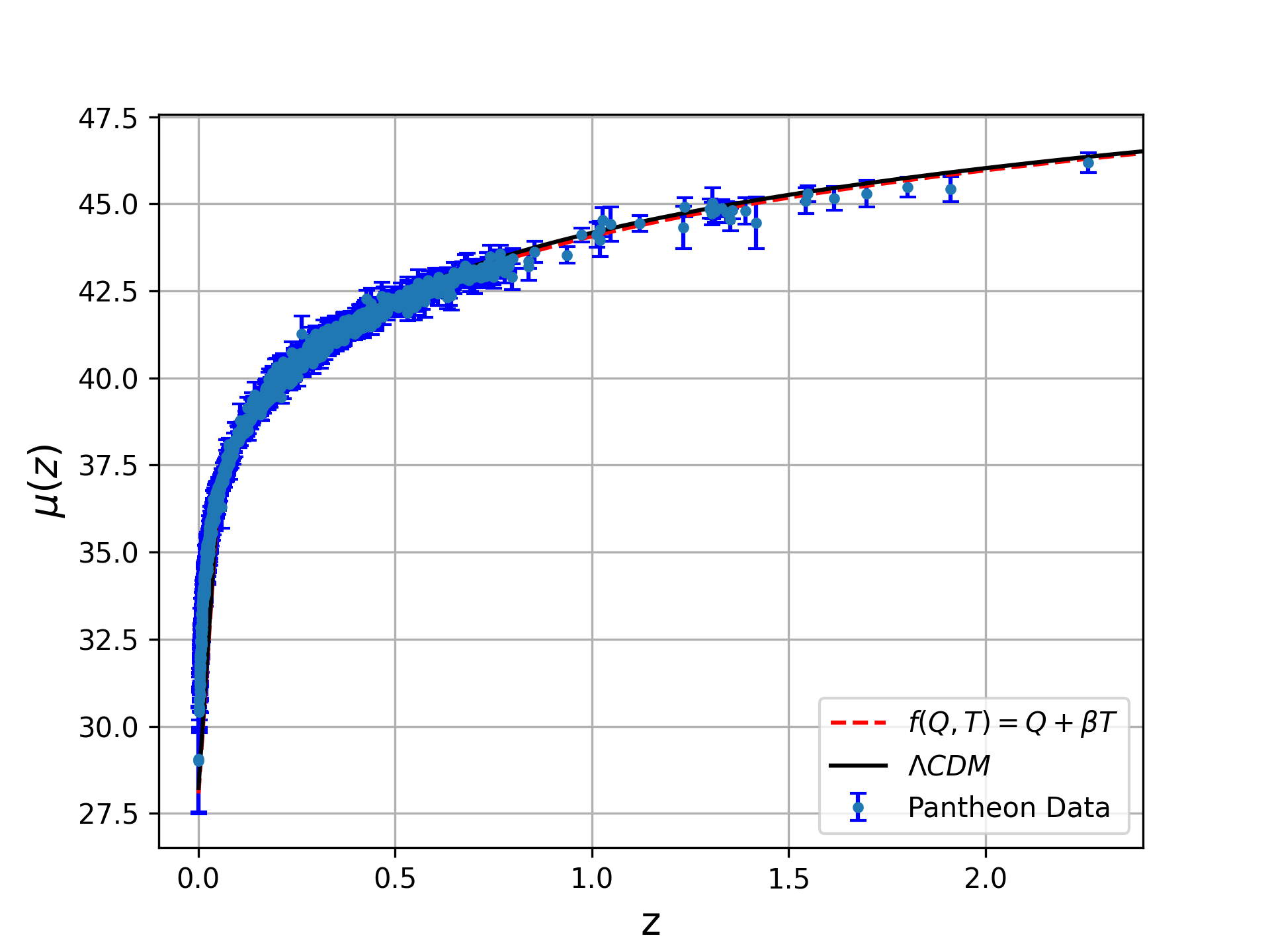}
        \caption{The plot of distance modulus $\mu (z)$ vs. redshift $z$ for Model I shown in red dotted line and $\Lambda CDM$ in black solid line which shows an excellent fit with the 1701 points of Pantheon datasets\cite{brout2022pantheon+} shown with it's error bars.}
        \label{distance modulus(Pantheon_data,n=0)}
    \end{minipage}
    \hspace{0.05\textwidth}
    \begin{minipage}[b]{0.46\textwidth}
        \centering
        \includegraphics[width=\textwidth]{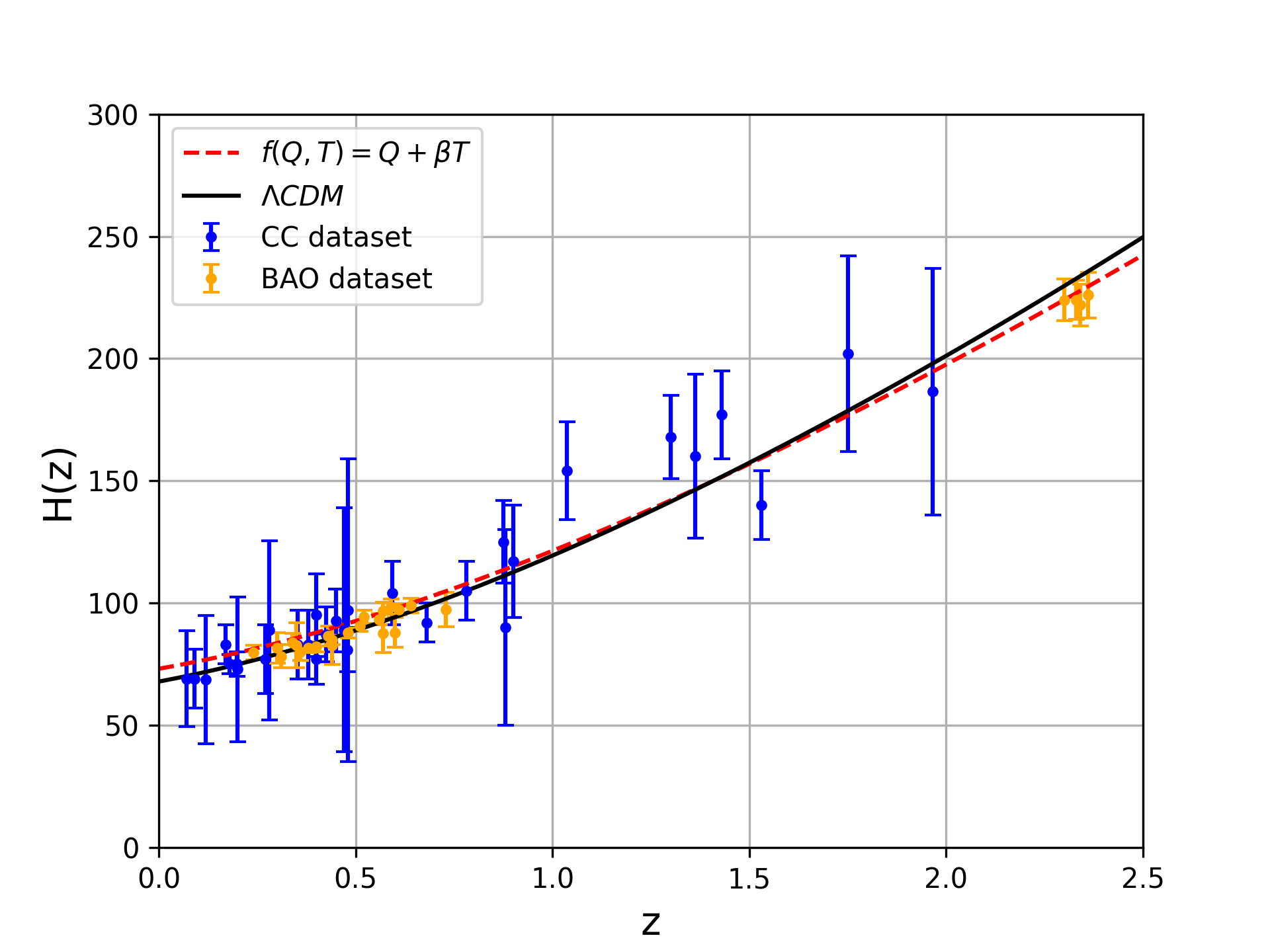}
        \caption{The plot of Hubble parameter $H(z)$ vs redshift $z$ for our model I, which is shown in red, and $\Lambda CDM$ which is shown in black line, shows an good match with the 57 points of the Hubble datasets\cite{yu2018hubble} shown with it's error bars.}
        \label{H(z)(Q+bT)}
    \end{minipage}
\end{figure}

\emph{Other Cosmological Behaviours:}

In order to further study the cosmological behaviour of this model,  we first examine the behaviour of the Hubble parameter w.r.t. redshift $z$. It is well known that the behaviour of the Hubble parameter in $\Lambda CDM$ cosmology is given by,
\begin{equation}\label{46}
    H_{\Lambda CDM}=H_{0}\sqrt{\Omega_{m0}(1+z)^{3}+1-\Omega_{m0}}
\end{equation}
where $H_{0}$ is the present value of the Hubble parameter and $\Omega_{m0}$ is the present value of matter density parameter defined as $\Omega_{m0}=\frac{\rho_{m0}}{3H_{0}^{2}}$ in Planck units. We set $H_{0}=67.8$ km/s/Mpc and $\Omega_{m0}=0.3$\cite{ade2016planck}.\\

\begin{figure}[htbp]
    \centering
    \begin{minipage}[b]{0.45\textwidth}
        \centering
        \includegraphics[width=\textwidth]{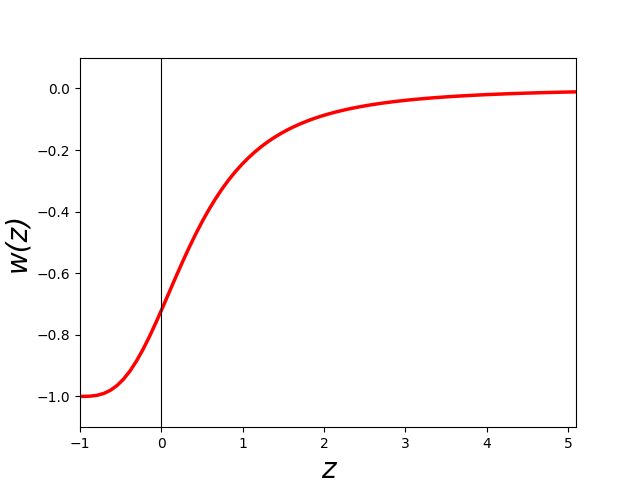}
        \caption{Variation of Equation of state parameter for the best fit value of $m=0.387$ from the MCMC analysis of four datasets}
        \label{w(z)(Q+bT)}
    \end{minipage}
    \hspace{0.05\textwidth}
    \begin{minipage}[b]{0.45\textwidth}
        \centering
        \includegraphics[width=\textwidth]{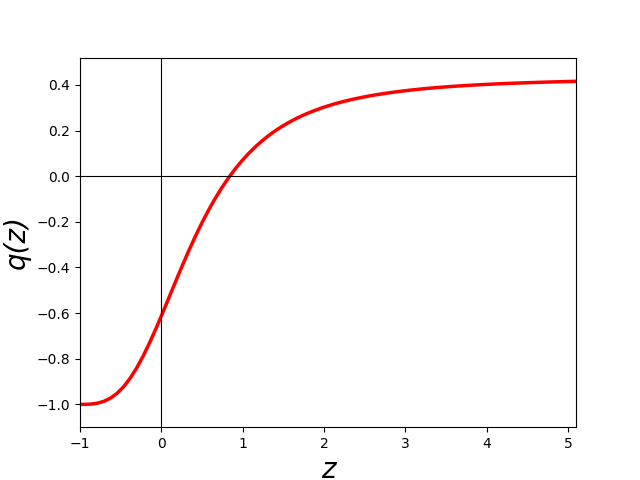}
        \caption{Variation of deceleration parameter $q(z)$ for the best fit value of $\beta=2.6$,$m=0.387$ from the joint analysis of four datasets}
        \label{q(z)(Q+bT)}
    \end{minipage}
\end{figure}

In Fig.\ref{H(z)(Q+bT)} we present the evolution of Hubble function Eq.(\ref{45}) with respect to redshift $z$ along with BAO and CC dataset. Here we have plotted the behaviour of $H(z)$ as obtained from Model I using the the best fit parameter values as shown in Fig.\ref{n=0_(4_legended_plot)} along with the $\Lambda CDM$ model and the observed data accompanied by error bars. The agreement between the model prediction and observed data is evident from the consistency of the error bars. This visual representation confirms the accuracy of our model in capturing the observed behaviour of the Hubble function.\\
The equation of state (EoS) parameter characterizes the relationship between energy density and pressure in the universe. Key cosmological phases are defined by specific values of the EoS parameter: $\omega=0$, the universe is in the dust phase; $\omega=1/3$ corresponds to the radiation-dominated phase; and $\omega=-1$ represents the vacuum energy, aligning with the $\Lambda CDM$ model. Recent discussions in cosmology focus on accelerating phases of the universe, which occur when $\omega < -1/3$ which includes the quintessence $(-1 < \omega < 0)$ and phantom phase ($\omega < -1$).\\
In Fig.\ref{w(z)(Q+bT)} we have shown the behavior of EoS parameter for the best fit parameter value of $m$ as obtained for Model I. The value of EoS parameter at $z=0$ is $\omega_{0}=-0.72_{-0.05}^{+0.03}$ which indicates an accelerating phase. Future behaviour indicates the EoS tending towards the value $-1$. As we go to higher redshifts, the EoS approaches $\omega=0$ which represented the dust phase.\\
Now, the deceleration parameter as a function of Hubble parameter $H$ is defined as,
\begin{equation}\label{47}
    q=-1-\frac{\dot{H}}{H^2}
\end{equation}\label{47.1}
The above equation w.r.t. redshift can be written as:
\begin{equation}
    q=-1-(1+z)\frac{1}{H(z)}\frac{dH}{dz}
\end{equation}
In cosmological models, the deceleration parameter $q$ is crucial in characterizing the evolution of the universe. It indicates whether the universe is in a decelerated phase($q>0$) or an accelerated phase($q<0$) of the universe. For Model I, the deceleration parameter $q$ is given by the expression:
\begin{equation}\label{48}
    q=-1-\frac{3m(8\pi+\beta)(1+z)^{3}}{[\beta+(16\pi+3\beta)(1+m(1+z)^{3})]}
\end{equation}
The behaviour of $q(z)$ is shown in Fig.\ref{q(z)(Q+bT)} according to the estimated best fit values of the model parameters $H_0$,$\beta$ and $m$ by MCMC. We find that there is a well behaved transition from deceleration to acceleration phase.
Also, we have found out that the present day value of deceleration parameter is, $q_{0}=-0.61_{-0.03}^{+0.03}$ which is negative at present time that represents the accelerating phase of the universe. The deceleration parameter tends towards $-1$ as we go to future redshifts.

\subsubsection{Model II : $f(Q,T)=Q^{2}+\beta T$}
\label{model II}
For the second example, we will consider function $f(Q,T)$ has the simple form $f(Q,T)=Q^{2}+\beta T$, where $\beta$ is constant. Then we obtain,
\begin{equation}\label{49}
F=f_{Q}=2 Q\; ; \;8\pi \tilde{G}=f_{T}=\beta
\end{equation}

Similar to Model I, solving for $p$ and $\rho$ from the Friedmann equations (\ref{10}) and (\ref{11}) for this particular form, we obtain the equation of state parameter as,
\begin{equation}\label{50}
    \omega=\frac{(16\pi+3\beta)\dot{H}+\frac{3}{2}(8\pi+\beta)H^{2}}{\beta \dot{H}-\frac{3}{2}(8\pi+\beta)H^{2}}
\end{equation}
Like before, using Eq.(\ref{16}) and Eq.(\ref{40}) we get the differential equation for the Hubble parameter as, 
\begin{equation}\label{51}
    \frac{dH}{dz}=\frac{3m(8\pi+\beta)(1+z)^{2}H(z)}{2[\beta+(16\pi+3\beta)(1+m(1+z)^{3})]}
\end{equation}
Solving the above equation we finally obtain the solution for the Hubble parameter as,
\begin{equation}\label{52}
    H(z)=H_{0}\left(\frac{\beta + (16\pi+3\beta)(1+m(1+z)^{3})}{\beta+(16\pi+3\beta)(1+m)}\right)^{l}
\end{equation}
where $l=\frac{(8\pi+\beta)}{2(16\pi+3\beta)}$ and $H_{0}$ is Hubble parameter value at present day, i.e at $z=0$.
Therefore, like Model I, the parameters that we want to constrain are $(H_0,\ \beta,\ m)$\\

\begin{figure}[h!]
\centering
\includegraphics[scale=0.6]{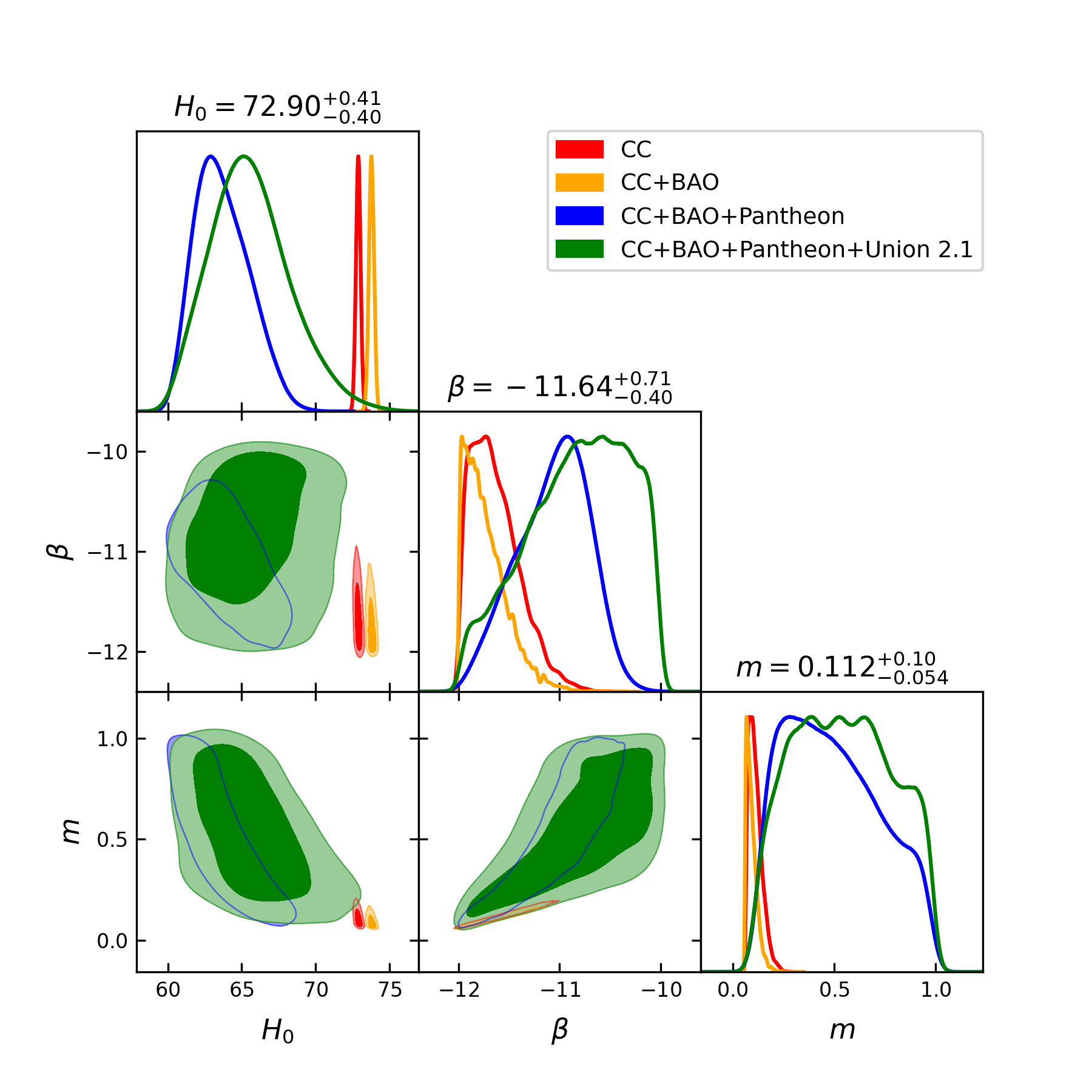}
\caption{\label{n=1_(4_legended_plot)}Joint likelihood contours of the model parameters ($H_0$,$\beta$,$m$) of Model II. The deeper shade show 68\% credible level (C.L.) and the lighter shade show 95\% credible level (C.L.)} 
\end{figure}

Fig \ref{n=1_(4_legended_plot)} contour plot show the the best fit values for the parameters of this model. The best fit value of the model parameters are shown in Table \ref{Table_2} for different datasets and the joint analysis along with the priors. As we can see joint analysis of CC+BAO+Pantheon gives the present value of $H_0$ as $73.79_{-0.47}^{+0.45}$ km/s/Mpc, thus alleviating the Hubble tension. Including the Union data, reduces the present value of $H_0$ to $72.90_{-0.40}^{+0.41}$ km/s/Mpc, which still fares much better than $\Lambda$CDM, thus reducing the tension with SHOES data.   In Fig.\ref{diatance modulus(Pantheon_data,n=1)} we have plotted our model and $\Lambda CDM$ model with the best fit parameter values obtained from MCMC. The plot also displays the Pantheon+ SNe Ia findings, 1701 data points, with their error bars, allowing for a direct comparison of the two models. As we can see our model fits the data as good as $\Lambda$CDM.

\begin{table*}[h!]
\caption{\label{Table_2}Best fit parameter values for Model II from MCMC}
\vspace{2mm}
\begin{tabular}{m{5cm} m{3.5cm} m{2.5cm} m{2cm}}
\hline
 Datasets & $H_{0}(km/s/Mpc)$ & $\beta$ & $m$ \\ 
\hline 
\vspace{1mm}
Priors & (60,80) & (-12,-10) & (0,1)\\ 
\vspace{1mm}
 CC & $65.7.19_{-5.8}^{+7.3}$ & $-10.82_{-1.1}^{+0.84}$ & $0.54_{-0.45}^{+0.47}$ \\
 \vspace{1mm}
 CC+BAO & $63.8_{-3.9}^{+5.2}$ & $-11.09_{-0.87}^{+0.82}$ & $0.49_{-0.40}^{+0.51}$ \\
 \vspace{1mm}
 CC+BAO+Pantheon & $73.79_{-0.47}^{+0.45}$ & $-11.72_{-0.30}^{+0.74}$ & $0.094_{-0.040}^{+0.091}$ \\
 \vspace{1mm}
CC+BAO+Pantheon+Union & $72.90_{-0.40}^{+0.41}$ & $-11.64_{-0.40}^{+0.71}$ & $0.112_{-0.054}^{+0.1}$ \\
\hline
\end{tabular}
\end{table*}

\begin{figure}[htbp]
    \centering
    \begin{minipage}[b]{0.46\textwidth}
        \centering
        \includegraphics[width=\textwidth]{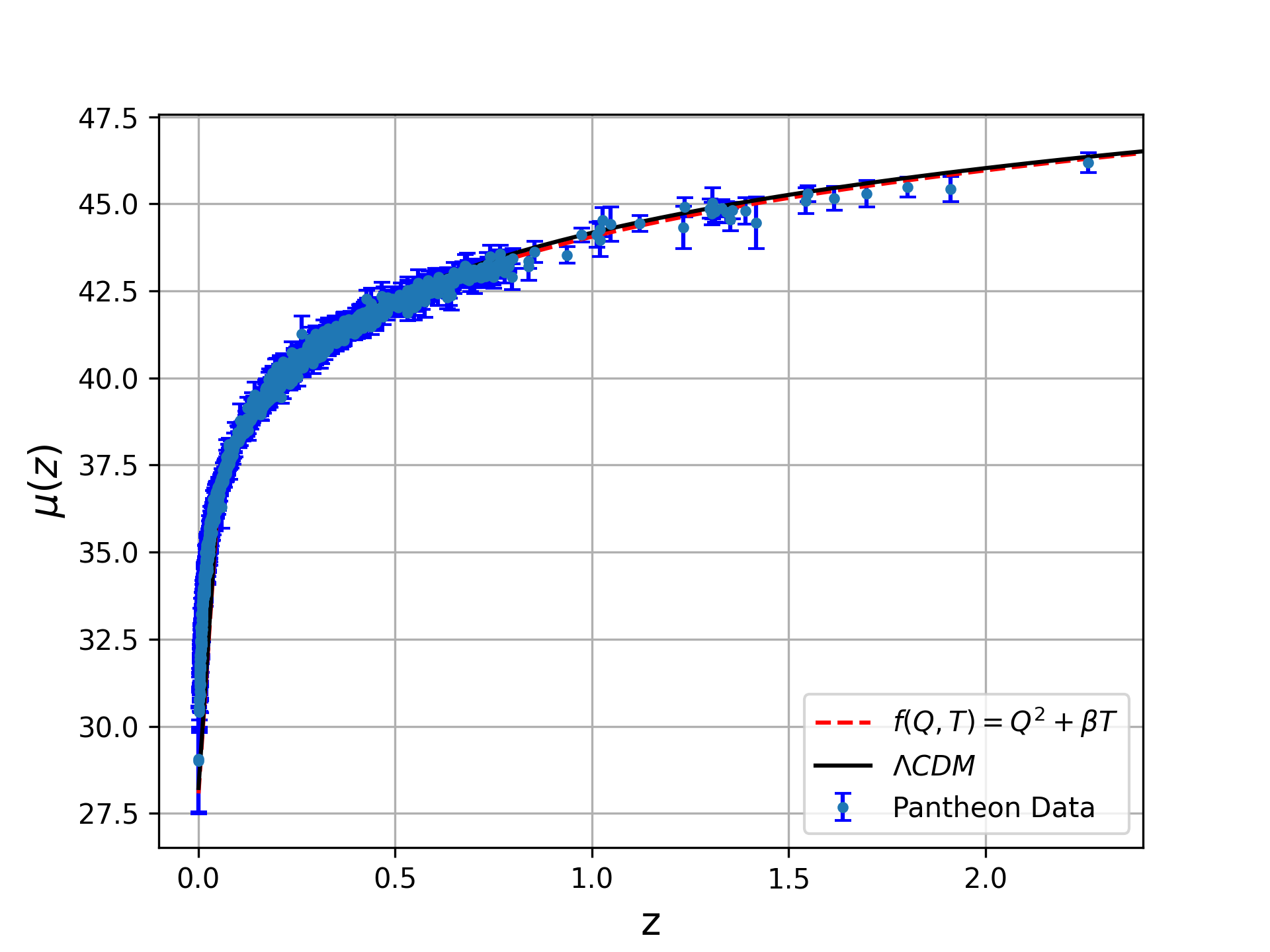}
        \caption{The plot of distance modulus $\mu(z)$ vs. redshift $z$ along with the corresponding error bars\cite{brout2022pantheon+} and the best-fit theoretical model for distance modulus function for Model II . $\Lambda CDM$ model also is shown in black solid line. }
        \label{diatance modulus(Pantheon_data,n=1)}
    \end{minipage}
    \hspace{0.05\textwidth}
    \begin{minipage}[b]{0.46\textwidth}
        \centering
        \includegraphics[width=\textwidth]{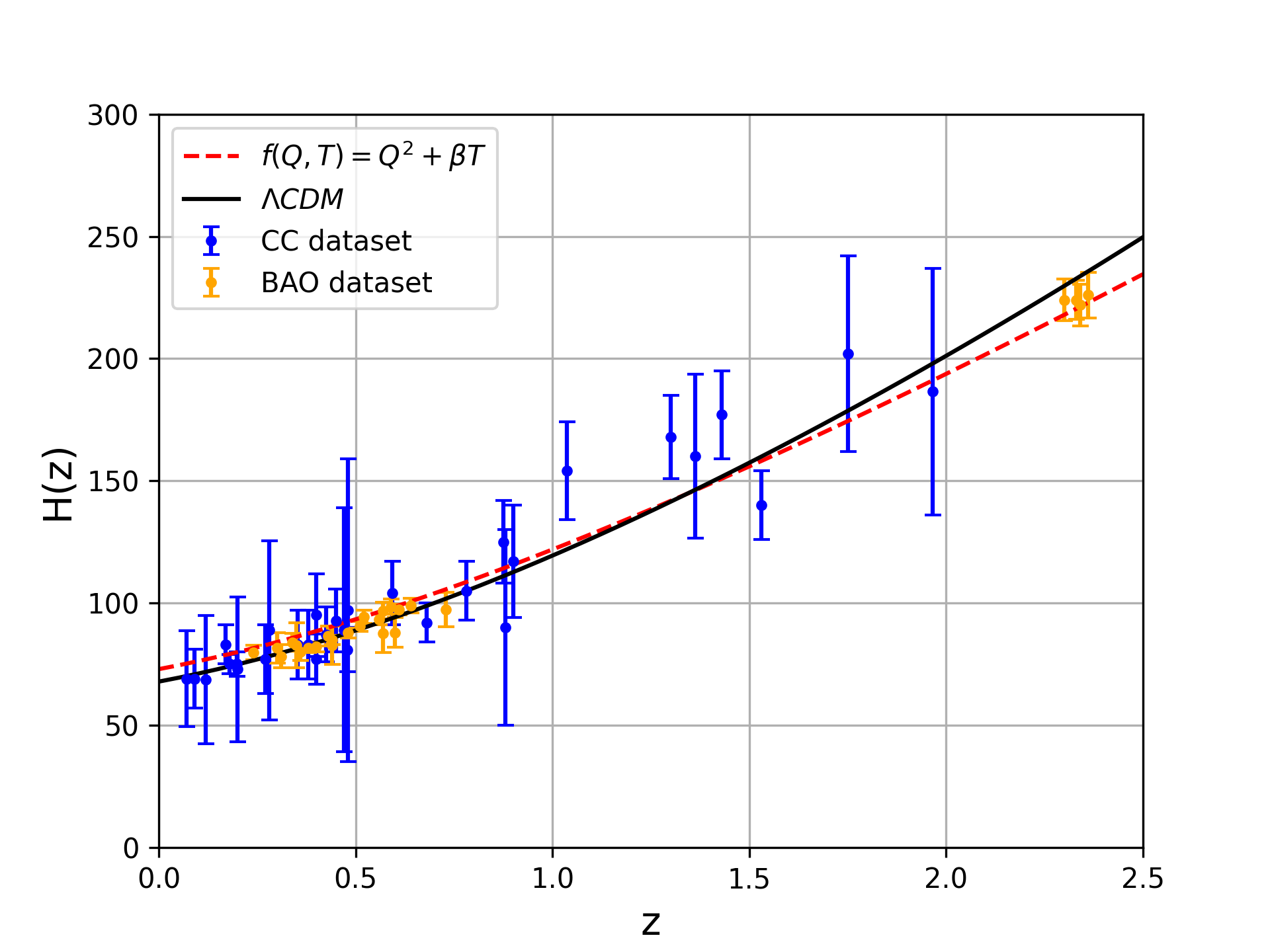}
        \caption{Comparing observed CC and BAO data with our theoretical model II for the best-fit parameter values. Our model is shown in red and $\Lambda CDM$ which is shown in black line, shows an good match with the 57 points of the Hubble dataset\cite{yu2018hubble}.}
        \label{H(z)(Q2+bT)}
    \end{minipage}
\end{figure}
\begin{figure}[htbp]
    \centering
    \begin{minipage}[b]{0.45\textwidth}
        \centering
        \includegraphics[width=\textwidth]{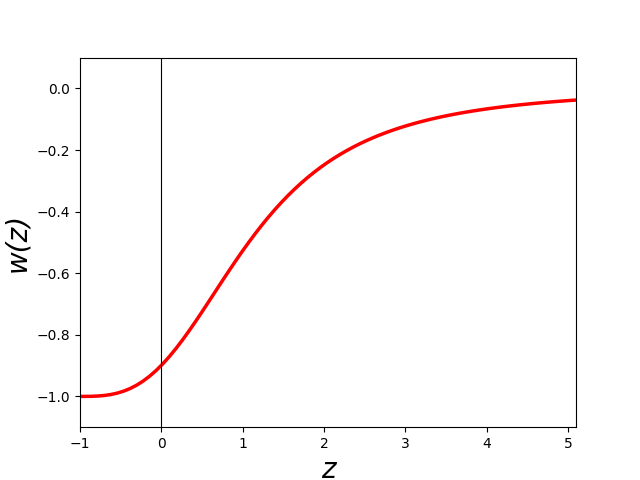}
        \caption{Variation of Equation of state parameter for the best fit value of $m=0.112$ from the MCMC analysis of four datasets}
        \label{w(z)(Q2+bT)}
    \end{minipage}
    \hspace{0.05\textwidth}
    \begin{minipage}[b]{0.45\textwidth}
        \centering
        \includegraphics[width=\textwidth]{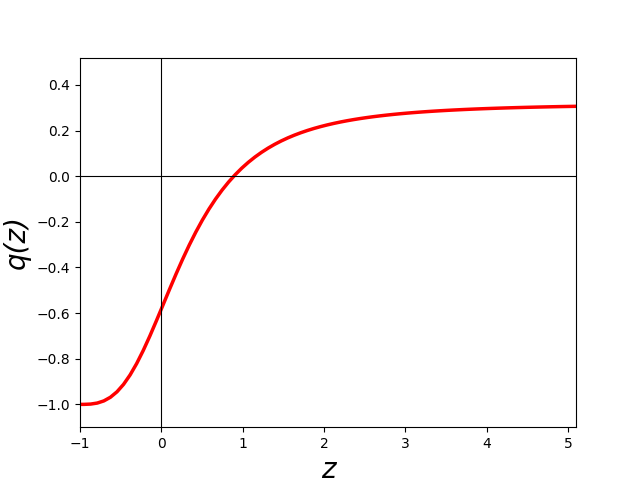}
        \caption{Variation of deceleration parameter $q(z)$ for the best fit value of $\beta=-11.64$,$m=0.112$ from the joint analysis of four datasets}
        \label{q(z)(Q2+bT)}
    \end{minipage}
\end{figure}
\emph{Other Cosmological Behaviours:}

Similar to Model I, we study the cosmological behaviour of Model II. Like the earlier case, in Fig.\ref{H(z)(Q2+bT)} we present the evolution of Hubble function Eq.(\ref{52}) with respect to redshift $z$ by using the best fit parameter values shown in Fig.\ref{n=1_(4_legended_plot)} for Model II  along with the $\Lambda CDM$ model and the observed CC and BAO data, accompanied by error bars. The agreement between the model prediction and observed data is as good as the earlier case.\\
Similar to the Model I, we have shown the behavior of EoS parameter in Fig.\ref{w(z)(Q2+bT)} for the best fit parameter value of $m$ obtained for Model II. The value of EoS parameter at present day($z=0$) is $\omega_{0}=-0.89_{-0.05}^{+0.07}$ which indicates an accelerating phase of the universe. 

The equation of deceleration parameter $q$ takes
the form for our model reads as,
\begin{equation}\label{53}
    q=-1-\frac{3m(8\pi+\beta)(1+z)^{3}}{2[\beta+(16\pi+3\beta)(1+m(1+z)^{3})]}
\end{equation}
The behavior of deceleration parameter is shown in Fig.\ref{q(z)(Q2+bT)} for the best fit parameter values of $\beta$ and $m$. We found that there is a well behaved transition from deceleration to acceleration phase and the present day value of $q$ is $q_{0}=-0.58_{-0.03}^{+0.02}$ which is negative at present time indicating the accelerating phase of the universe.
Like Model I, here also both $\omega$ and $q$ tend towards $-1$ for future redshifts.

\subsubsection{Model III : $f(Q,T)=Q^{3}+\beta T$}
\label{model III}
For the third example, we consider cubic dependence on $Q$. The function $f(Q,T)$ has the form $f(Q,T)=Q^{3}+\beta T$, where $\beta$ is constant. Here,
\begin{equation}\label{54}
F=f_{Q}=3Q^{2}\; ; \;8\pi \tilde{G}=f_{T}=\beta
\end{equation}
Similar to the earlier cases, from the Friedmann equations (\ref{10}) and (\ref{11}) for this particular form, we obtain the form of $\omega$ as,
\begin{equation}\label{55}
    \omega=\frac{(16\pi+3\beta)\dot{H}+(8\pi+\beta)H^{2}}{\beta \dot{H}-(8\pi+\beta)H^{2}}
\end{equation}
By using Eq.(\ref{16}) and Eq.(\ref{40}) we obtain the differential equation for Hubble parameter as, 
\begin{equation}\label{56}
    \frac{dH}{dz}=\frac{m(8\pi+\beta)(1+z)^{2}H(z)}{[\beta+(16\pi+3\beta)(1+m(1+z)^{3})]}
\end{equation}
Finally, solving the equation we obtain the Hubble parameter as,
\begin{equation}\label{57}
    H(z)=H_{0}\left(\frac{\beta + (16\pi+3\beta)(1+m(1+z)^{3})}{\beta+(16\pi+3\beta)(1+m)}\right)^{l}
\end{equation}
where $l=\frac{(8\pi+\beta)}{3(16\pi+3\beta)}$ and $H_{0}$ is Hubble parameter value at present day, i.e at $z=0$.\\

\begin{figure}[h!]
\centering
\includegraphics[scale=0.6]{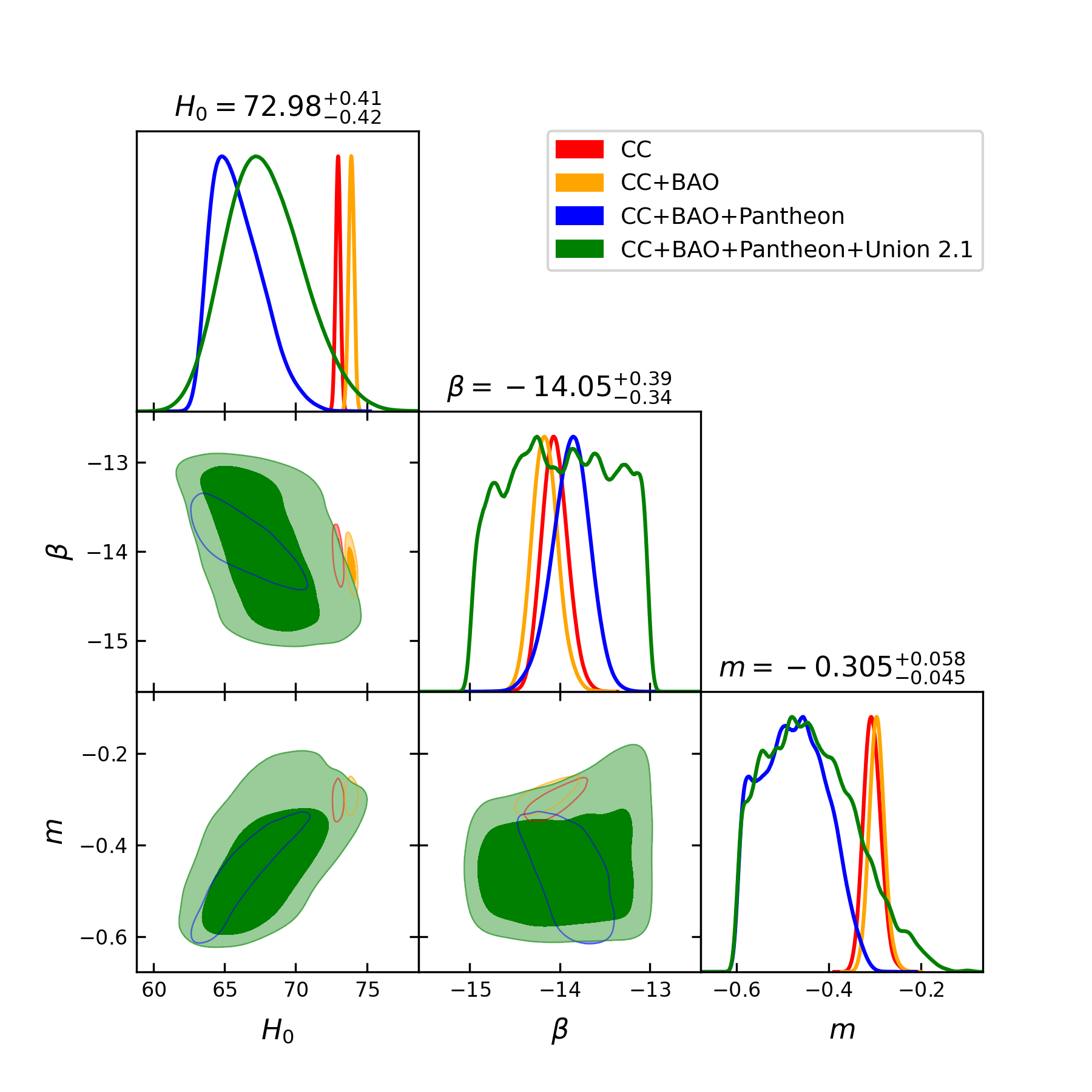}
\caption{\label{n=2_(4_legended_plot)}Joint likelihood contours of the model parameters ($H_0$,$\beta$,$m$) of Model III. The contours are at 68\%, 95\% credible level (C.L.)} 
\end{figure}
\begin{table*}[h!]
\caption{\label{Table_3}Best fit parameter values for Model III from MCMC}
\vspace{2mm}
\begin{tabular}{m{5cm} m{3.5cm} m{2.5cm} m{2cm}}
\hline
 Datasets & $H_{0}(km/s/Mpc)$ & $\beta$ & $m$ \\ 
\hline 
\vspace{1mm}
Priors & (60,80) & (-15,-13)& (-0.6,0)\\ 
\vspace{1mm}
 CC & $68_{-6}^{+8}$ & $-14_{-0.99}^{+0.99}$ & $-0.44_{-0.17}^{+0.26}$ \\
 \vspace{1mm}
 CC+BAO & $66_{-3.4}^{+5}$ & $-13.88_{-0.60}^{+0.54}$ & $-0.47_{-0.13}^{+0.15}$ \\
 \vspace{1mm}
 CC+BAO+Pantheon & $73.92_{-0.49}^{+0.48}$ & $-14.17_{-0.36}^{+0.41}$ & $-0.295_{-0.041}^{+0.050}$ \\
 \vspace{1mm}
CC+BAO+Pantheon+Union & $72.98_{-0.42}^{+0.41}$ & $-14.05_{-0.34}^{+0.39}$ & $-0.305_{-0.045}^{+0.058}$ \\
\hline
\end{tabular}
\end{table*}

Fig \ref{n=2_(4_legended_plot)} contour plot show the the best fit values for the parameters of this model. The best fit value of the model parameters are shown in Table \ref{Table_3} for different datasets and the joint analysis. In terms of Hubble tension, this model performs as good as Model II. The present value of Hubble parameter is nearly same as what we obtained for Model II. The best fit value of the model parameters ($\beta,\ m$) from the joint analysis of CC+BAO+Pantheon+Union data are ($-14.05_{-0.34}^{+0.39},\ -0.305_{-0.045}^{+0.058} $). Unlike the earlier cases, the best fit value of $m$ turns out to be negative for Model III. Similar to the earlier cases, in Fig.\ref{diatance modulus(Pantheon_data,n=2)} we have plotted our model and $\Lambda CDM$ model with the best fit parameter values obtained from MCMC along with Pantheon+SNe Ia data.

\begin{figure}[htbp]
    \centering
    \begin{minipage}[b]{0.46\textwidth}
        \centering
        \includegraphics[width=\textwidth]{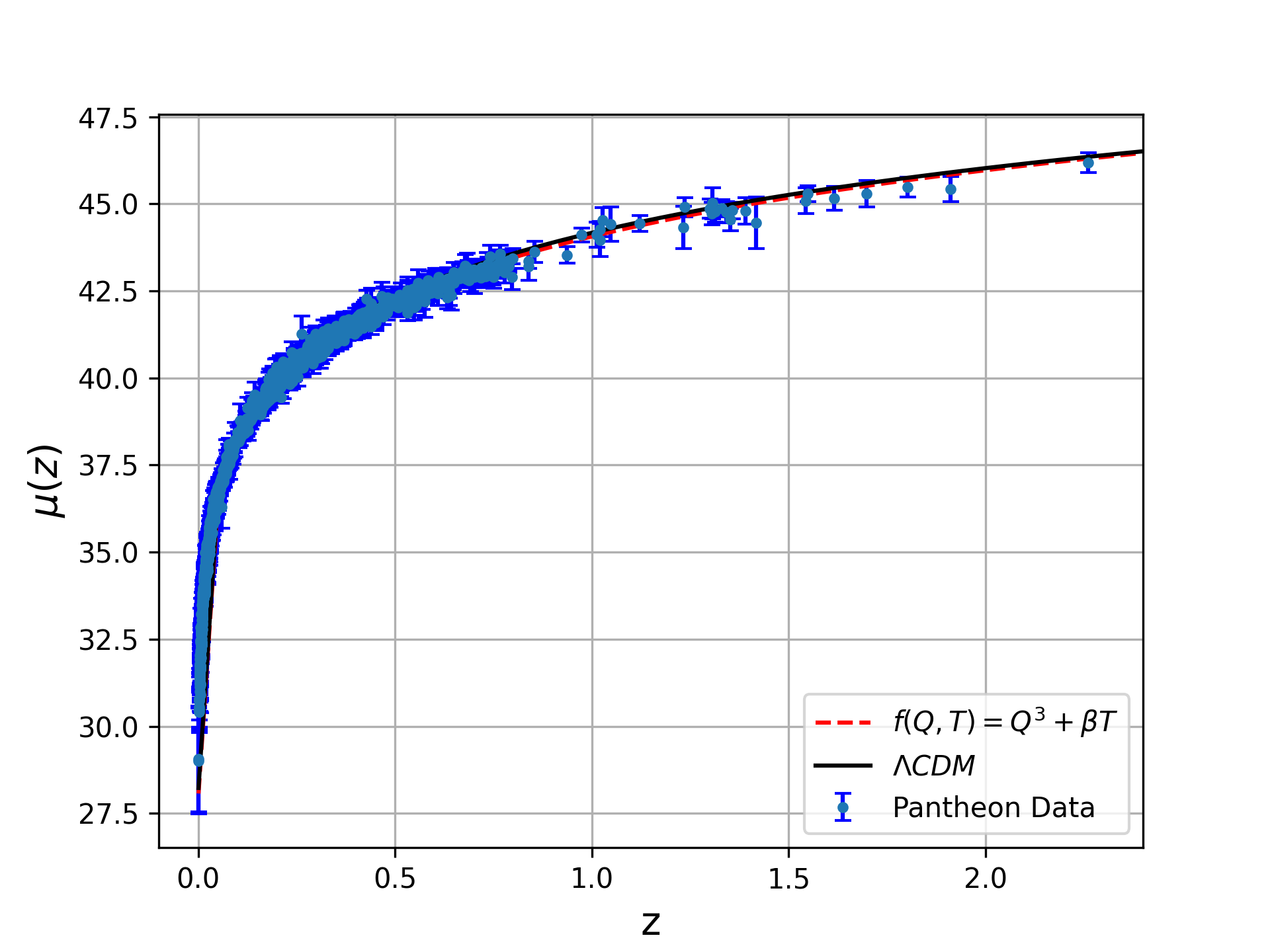}
        \caption{The plot of distance modulus $\mu(z)$ vs. redshift $z$ for Model III shown in red dotted line and $\Lambda CDM$ in black solid line which shows an excellent match with the 1701 data points\cite{brout2022pantheon+} of Pantheon dataset shown with it’s error bars.}
        \label{diatance modulus(Pantheon_data,n=2)}
    \end{minipage}
    \hspace{0.05\textwidth}
    \begin{minipage}[b]{0.46\textwidth}
        \centering
        \includegraphics[width=\textwidth]{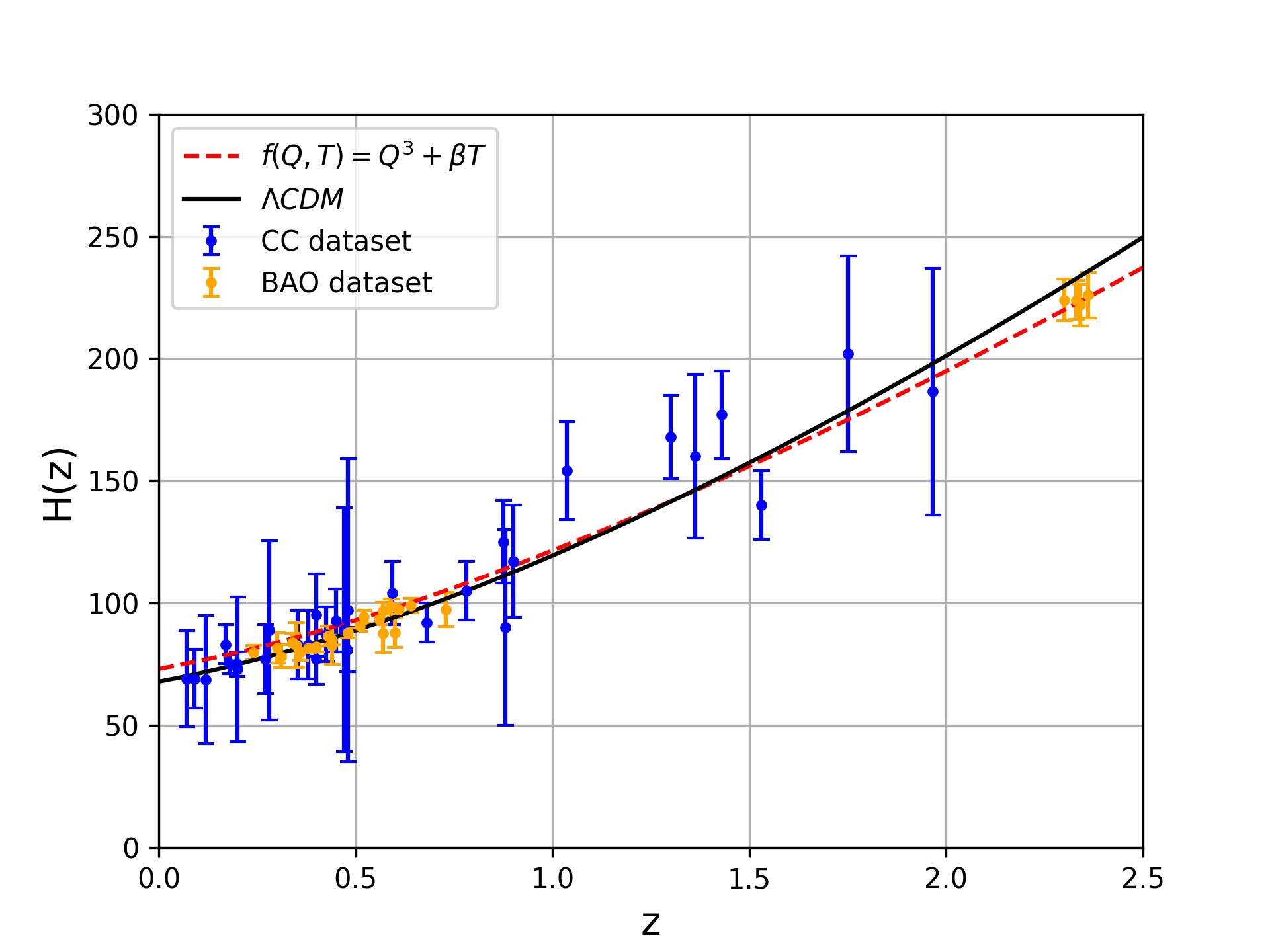}
        \caption{The plot of Hubble parameter $H(z)$ vs redshift $z$. The black solid line corresponds to the $\Lambda CDM$ model and  the red dotted line represent our model III, shows an good match with the 57 points of the Hubble data\cite{yu2018hubble}.}
        \label{H(z)(Q3+bT)}
    \end{minipage}
\end{figure}

\emph{Other Cosmological Behaviours:}

\begin{figure}[htbp]
    \centering
    \begin{minipage}[b]{0.45\textwidth}
        \centering
        \includegraphics[width=\textwidth]{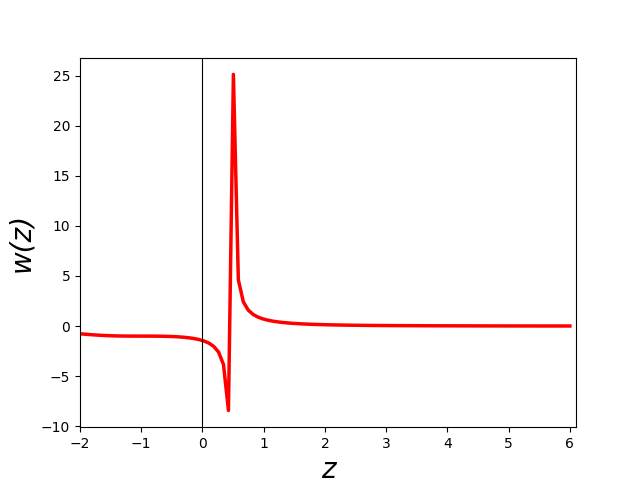}
        \caption{Variation of Equation of state parameter for the best fit value of $m=-0.305$ from the MCMC analysis of four datasets}
        \label{w(z)(Q3+bT)}
    \end{minipage}
    \hspace{0.05\textwidth}
    \begin{minipage}[b]{0.45\textwidth}
        \centering
        \includegraphics[width=\textwidth]{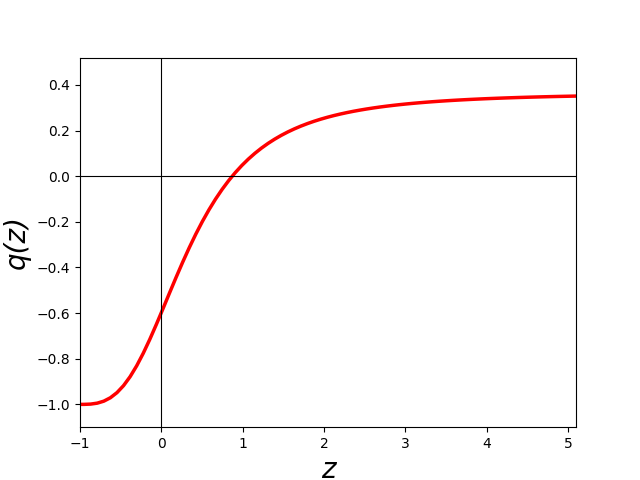}
        \caption{Variation of deceleration parameter $q(z)$ for the best fit value of $\beta=-14.05$,$m=-0.305$ from the joint analysis of four datasets}
        \label{q(z)(Q3+bT)}
    \end{minipage}
\end{figure}

In Fig.\ref{H(z)(Q3+bT)} we present the evolution of Hubble function Eq.(\ref{57}) with respect to redshift $z$. By using the best fit parameter value shown in Fig.\ref{n=2_(4_legended_plot)}  along with the $\Lambda CDM$ model and the observed data on $H(z)$, accompanied by error bars. The agreement between the model prediction and observed data is again evident from the consistency of the error bars. \\
Similar to the Model I, we have shown the behavior of EoS parameter in Fig.\ref{w(z)(Q3+bT)} for the best fit parameter value $m$. The value of EoS parameter at present day($z=0$) is $\omega_{0}=-1.44_{-0.1}^{+0.11}$ which indicates an accelerating phase of the universe. However we see that the EoS makes a sudden transition from negative to positive value in and around $z=1$, beyond which it tends toward $\omega=0$. Therefore, we can conclude that a negative value of $m$ fails to give the correct behaviour of the EoS and hence is not acceptable, thereby disfavouring this model over the others.

The equation of deceleration parameter $q$ takes
the form for our model reads as,
\begin{equation}\label{58}
    q=-1-\frac{m(8\pi+\beta)(1+z)^{3}}{[\beta+(16\pi+3\beta)(1+m(1+z)^{3})]}
\end{equation}
The behavior of deceleration parameter is shown in Fig.\ref{q(z)(Q3+bT)} for the best fit parameter values of $\beta$ and $m$. We found that there is a well behaved transition from deceleration to acceleration phase and the present day value of $q$ is $q_{0}=-0.59_{-0.03}^{+0.01}$ indicating the accelerating phase of the universe.

\subsection{Model IV : $f(Q,T)=-\alpha Q-\beta T^{2}$}
\label{model IV}
As a fourth example of cosmological model in $f(Q,T)$ gravity, we consider the case having non-linear dependence on $T$. We consider the following form $f(Q,T)=-\alpha Q-\beta T^{2}$, where $\alpha>0$ and $\beta>0$ are constants. For this model we have,
\begin{equation}\label{59}
F=f_{Q}=-\alpha
\end{equation}
\begin{equation}\label{60}
8\pi \tilde{G}=f_{T}=-2\beta T=2\beta(1-3\omega)\rho
\end{equation}
From Eq.(\ref{14}), the cosmological density will be,
\begin{equation}\label{61}
\rho=\frac{6\alpha H^{2}-\beta (3\omega-1)^{2}\rho^{2}}{16\pi[1+(1+\omega)\tilde{G}]}
\end{equation}
which has the physical solution,
\begin{equation}\label{62}
\rho=\frac{8\pi\left[\sqrt{1+3\alpha\beta(1-3\omega)(\omega+5) H^{2}/32\pi^{2}}-1\right]}{\beta(1-3\omega)(\omega+5)}    
\end{equation}
Now, the evolution equation of Hubble parameter for this model takes the form,
\begin{multline}\label{63}
\dot{H}=-\frac{32\pi^{2}(1+\omega)}{\alpha \beta(1-3\omega)(\omega+5)}\times 
\\
\left[\sqrt{1+3\alpha\beta(1-3\omega)(\omega+5) H^{2}/32\pi^{2}}-1\right]\times
\\
\left[1-\frac{2\beta(3\omega-1)\left(\sqrt{1+3\alpha \beta H^{2}(1-3\omega)(\omega+5)/32\pi^{2}}-1\right)}{\beta(1-3\omega)(\omega+5)}\right]
\end{multline}
The corresponding evolution of Hubble parameter with respect to redshift $z$ will be,
\begin{multline}\label{64}
(1+z)H(z)\frac{dH}{dz}=\frac{3(1+\omega)}{\delta}\left(\sqrt{1+\delta H^{2}}-1\right)\times
\\
\left(1-\frac{2\beta(3\omega-1)\left(\sqrt{1+\delta H^{2}}-1\right)}{\beta(1-3\omega)(\omega+5)}\right)    
\end{multline}
where, 
\begin{equation}\label{65}
\delta=\frac{3\alpha\beta(1-3\omega)(\omega+5)}{32\pi^{2}}
\end{equation}
and $\omega$ is given by the Eq.\ref{40}. The parameters that will be constrained by data for the present case are ($H_0,\ \alpha,\ \beta,\ m$).

\begin{figure*}[h!]
\centering
\includegraphics[height=11cm,width=12cm]{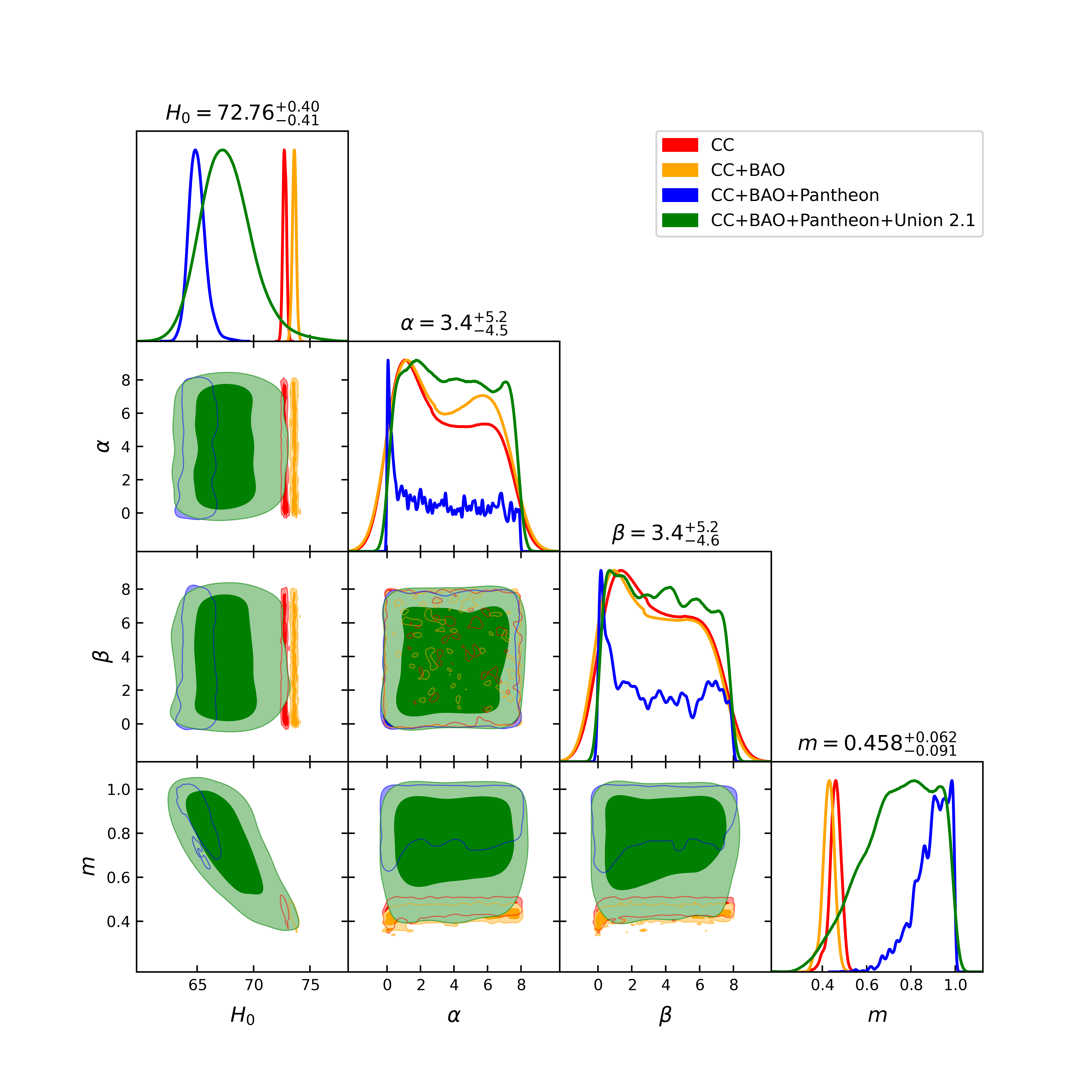}
\caption{\label{-aQ-bT2_(4_egended+plot)}Posterior distribution of the model parameters listed in Table IV utilizing CC, BAO, SN Ia observations. The deeper shade show 68\% credible level (C.L.) and the lighter shade show 95\% credible level (C.L.)} 
\end{figure*}

Fig \ref{-aQ-bT2_(4_egended+plot)} contour plot show the the best fit values for the parameters of this model. The best fit value of the model parameters are shown in Table \ref{Table_4} for different datasets and the joint analysis with the priors. As we can see joint analysis of CC+BAO+Pantheon gives the present value of $H_0$ as $73.61_{-0.47}^{+0.47}$ km/s/Mpc, thus alleviating the Hubble tension. The corresponding best fit values of the model parameters ($\alpha,\ \beta,\ m$) are ($3.1_{-4.2}^{+5.3},\ 3_{-4}^{+5.2},\ 0.428_{-0.072}^{+0.069}$). Joint analysis with the Union data, reduces the present value of $H_0$ to $72.76_{-0.41}^{+0.40}$ km/s/Mpc, which is still a large improvement over $\Lambda CDM$, thus reducing the tension with SHOES data. The corresponding best fit values of the model parameters ($\alpha,\ \beta,\ m$) are ($3.4_{-4.5}^{+5.2},\ 3.4_{-4.6}^{+5.2},\ 0.458_{-0.091}^{+0.062}$). In Fig.\ref{distance modulus(Pantheon_data,-aQ-bT2)} our model with the best parameter values obtained from joint analysis of CC+BAO+Pantheon+Union and $\Lambda CDM$ model has been plotted along with Pantheon+ SNe Ia findings for comparison. Like for the other cases, our model fits the data as good as $\Lambda$CDM.\\

\begin{table*}[h!]
\caption{\label{Table_4}Best fit parameter values for Model IV from MCMC}
\vspace{2mm}
\begin{tabular}{m{5cm} m{3cm} m{2cm} m{2cm} m{1.7cm}}
\hline
 Datasets & $H_{0}(km/s/Mpc)$ &  $\alpha$ & $\beta$ & $m$ \\ 
\hline 
\vspace{1mm}
Priors & (60,80) & (0,8) & (0,8) & (0,1)\\
\vspace{1mm}
 CC & $67.7_{-5.3}^{+6.9}$ & $4_{-4}^{+4}$ & 
$3.9_{-3.9}^{+4}$  & $0.73_{-0.37}^{+0.28}$ \\
\vspace{1mm}
 CC+BAO & $65_{-1.6}^{+2.4}$ & $3.6_{-3.6}^{+4.4}$ & $3.7_{-3.7}^{+4.2}$ & $0.90_{-0.26}^{+0.11}$ \\
 \vspace{1mm}
 CC+BAO+Pantheon & $73.61_{-0.47}^{+0.47}$ & $3.1_{-4.2}^{+5.3}$ & $3_{-4}^{+5.2}$ &$0.428_{-0.072}^{+0.069}$ \\
 \vspace{1mm}
CC+BAO+Pantheon+Union & $72.76_{-0.41}^{+0.40}$ & $3.4_{-4.5}^{+5.2}$ & $3.4_{-4.6}^{+5.2}$ & $0.458_{-0.091}^{+0.062}$ \\
\hline
\end{tabular}

\end{table*}
\begin{figure}[htbp]
    \centering
    \begin{minipage}[b]{0.46\textwidth}
        \centering
        \includegraphics[width=\textwidth]{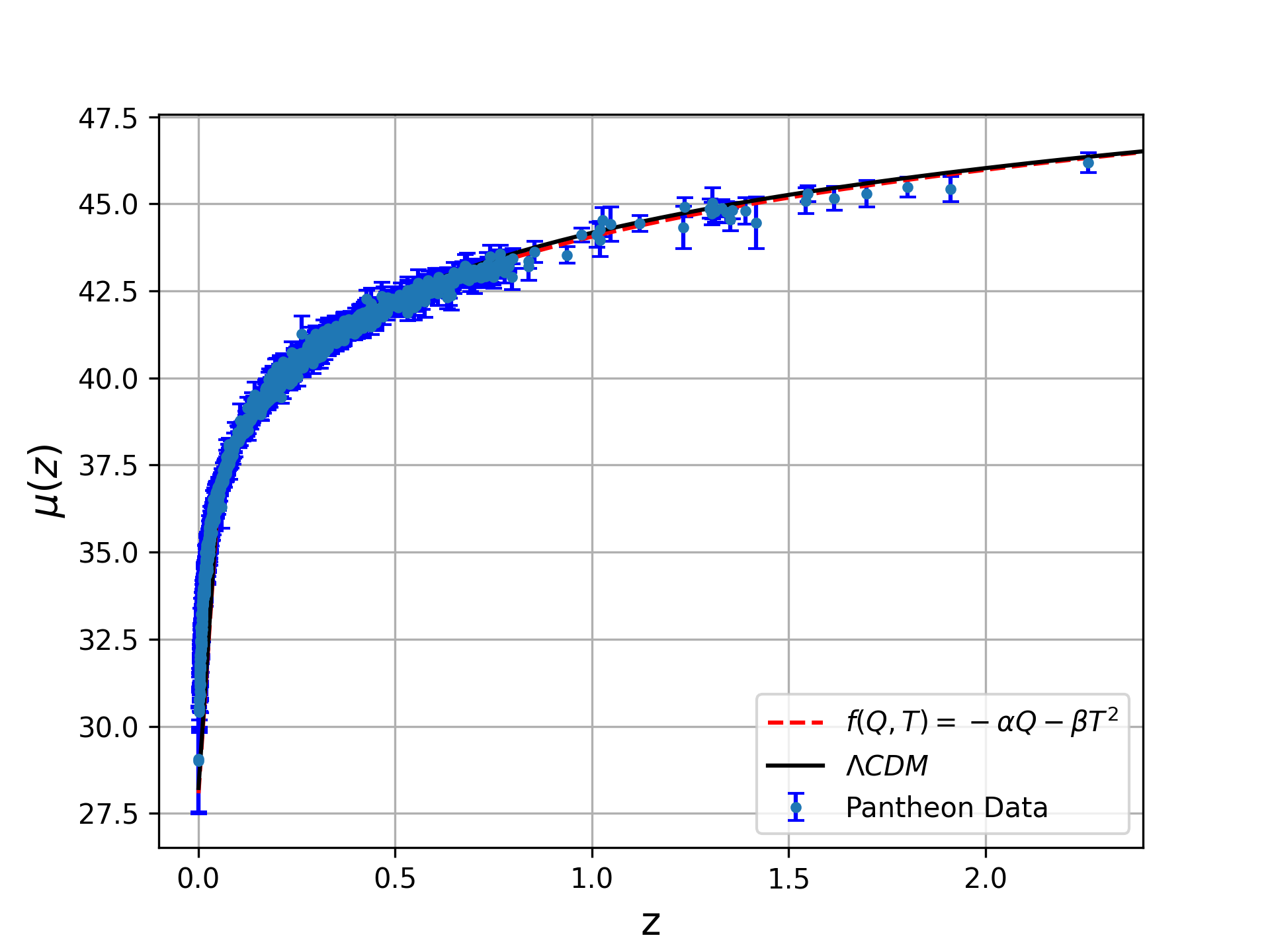}
        \caption{The plot of distance modulus $\mu(z)$ vs. redshift $z$ for Model IV shown in red dotted line and $\Lambda CDM$ in black solid line. Our model shows an excellent fit with the 1701 points\cite{brout2022pantheon+} of Pantheon dataset shown with it’s error bars.}
        \label{distance modulus(Pantheon_data,-aQ-bT2)}
    \end{minipage}
    \hspace{0.05\textwidth}
    \begin{minipage}[b]{0.46\textwidth}
        \centering
        \includegraphics[width=\textwidth]{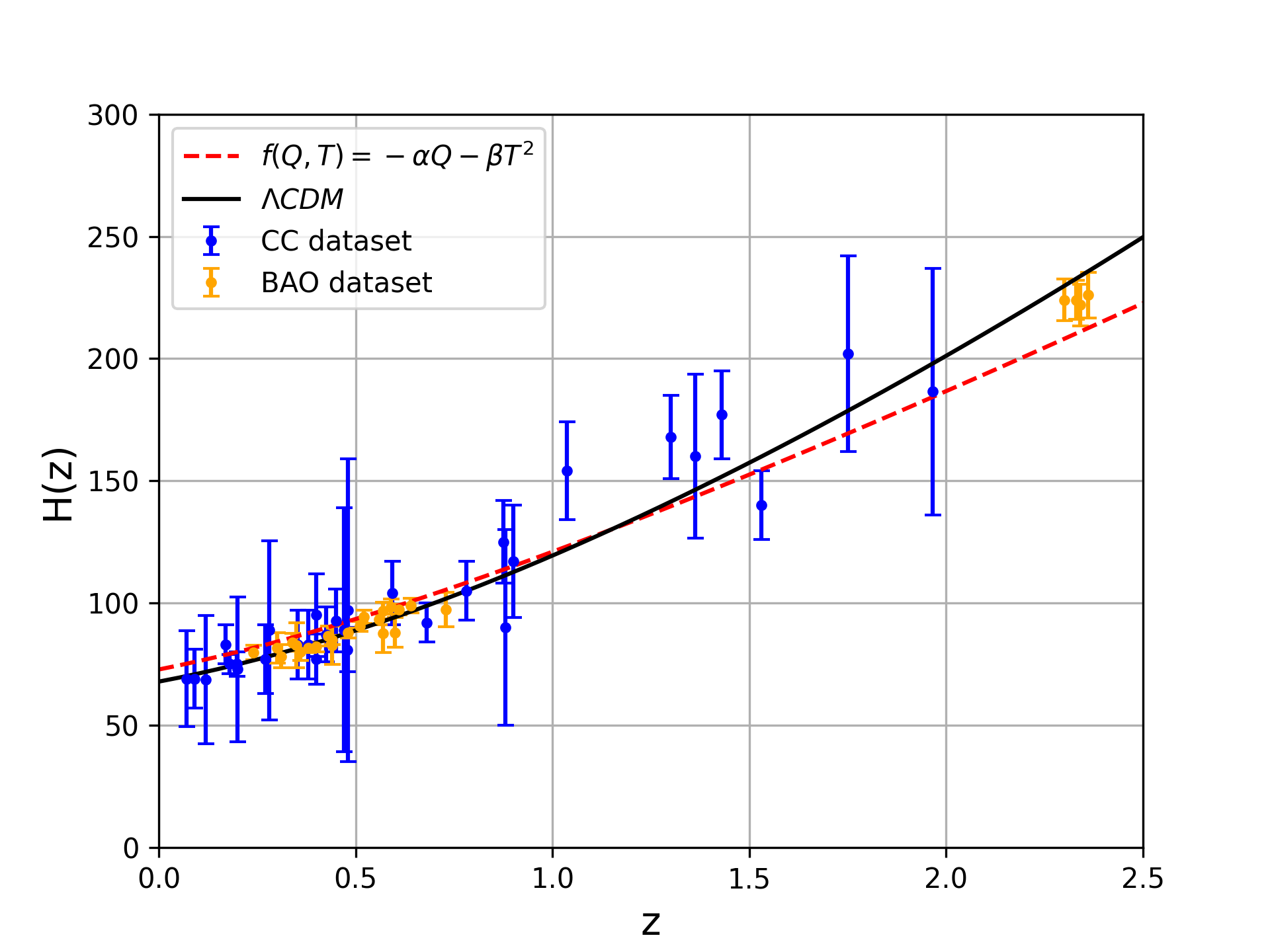}
        \caption{The plot of Hubble parameter $H(z)$ vs redshift $z$ for our model IV, which is shown in red, and $\Lambda CDM$ which is shown in black line. Our model shows an excellent match with the 57 points of the Hubble dataset\cite{yu2018hubble} shown with it's error bars.}
        \label{H(z)(-aQ-bT2)}
    \end{minipage}
\end{figure}
\emph{Other Cosmological Behaviours:}

Let us now proceed to the numerical investigation of this model. Similarly to the previous Model I, we set the same constants for the $\Lambda CDM$ model.\\
\begin{figure}[htbp]
    \centering
    \begin{minipage}[b]{0.45\textwidth}
        \centering
        \includegraphics[width=\textwidth]{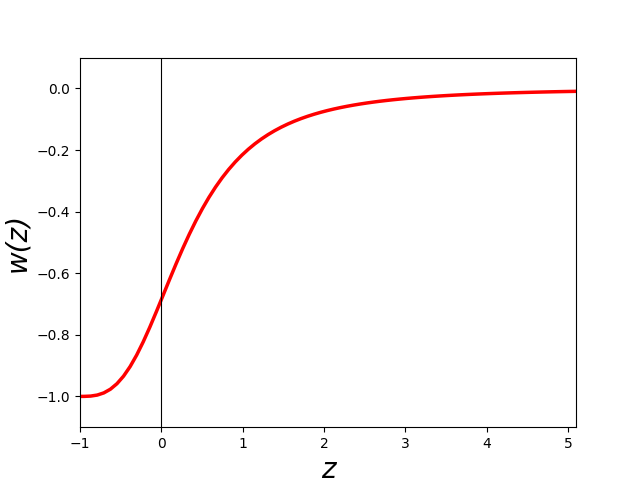}
        \caption{Variation of Equation of state parameter for the best fit value of $m=0.458$ from the MCMC analysis of four datasets}
        \label{w(z)(-aQ-bT2)}
    \end{minipage}
    \hspace{0.05\textwidth}
    \begin{minipage}[b]{0.45\textwidth}
        \centering
        \includegraphics[width=\textwidth]{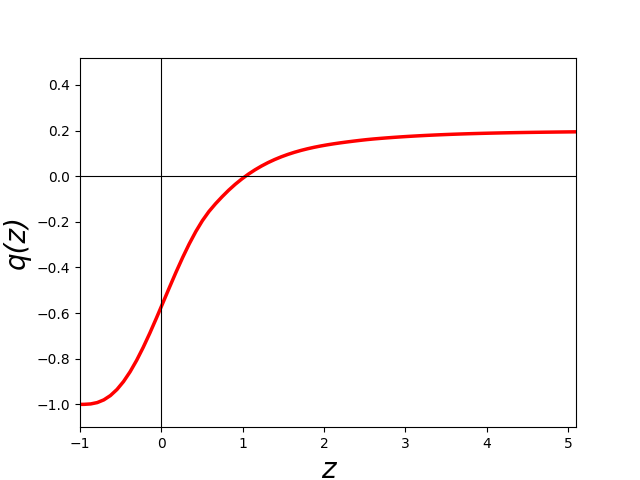}
        \caption{Variation of deceleration parameter $q(z)$ for the best fit value of $\alpha=3.4$,$\beta=3.4$,$m=0.458$ from the joint analysis of four datasets}
        \label{q(z)(-aQ-bT2)}
    \end{minipage}
\end{figure}
In Fig.\ref{H(z)(-aQ-bT2)} we present the evolution of Hubble function by solving the Eq.(\ref{64}) with respect to redshift $z$ by using the best fit parameter value shown in Fig.\ref{-aQ-bT2_(4_egended+plot)}  along with the $\Lambda CDM$ model and the observed data on $H(z)$, accompanied by error bars. Model IV agrees nearly as well with the observed data as $\Lambda$CDM (though the earlier models fair better). \\
Now, we have solved this model numerically and shown the behavior of EoS parameter in Fig.\ref{w(z)(-aQ-bT2)} for the best fit parameter value $m$. The value of EoS parameter at present day($z=0$) is $\omega_{0}=-0.68_{-0.05}^{+0.02}$ indicating, like for the other cases, an accelerating phase of the universe.

The behavior of deceleration parameter is shown in Fig.\ref{q(z)(-aQ-bT2)} for the best fit parameter values of $\alpha$,$\beta$ and $m$. The present day value of $q$ turns out to be $q_{0}=-0.54_{-0.05}^{+0.03}$.

\subsection{Model V : $f(Q,T)=Q^{-2}T^{2}$}
\label{model V}
For our fifth example of cosmological model in $f(Q,T)$ gravity, we will consider a pure non-minimally coupled case which has the form $f(Q,T)=Q^{-2}T^{2}$. Then we immediately obtain,
\begin{equation}\label{66}
F=f_{Q}=-(1/108H^{6})(3\omega-1)^{2}\rho^{2} 
\end{equation}
\begin{equation}\label{67}
    8\pi \tilde{G}=(1/18H^{4})(3\omega-1)\rho
\end{equation}
Then, the cosmological density will be,
\begin{equation}\label{68}
    \rho=\frac{-16\pi H^{4}}{[(1/9)(1+\omega)(3\omega-1)-(5/36)(3\omega-1)^{2}]}
\end{equation}
Now, the evolution of Hubble parameter for this model takes the form,
\begin{equation}\label{69}
    \dot{H}=-\frac{18(1+\omega)\left[8\pi + (1/18H^{4})(3\omega-1)\rho\right]H^{6}}{(3\omega-1)^{2}\rho}
\end{equation}
Now, the evolution of Hubble parameter with respect to redshift $z$ will be,
\begin{equation}\label{70}
    \frac{dH}{dz}=\frac{18(1+\omega)\left[8\pi + (1/18H^{4})(3\omega-1)\rho\right]H^{5}}{(1+z)(3\omega-1)^{2}\rho}
\end{equation}
The parameters that we will constrain for this case are ($H_0,\ m$).

\begin{figure}[h!]
\centering
\includegraphics[scale=0.55]{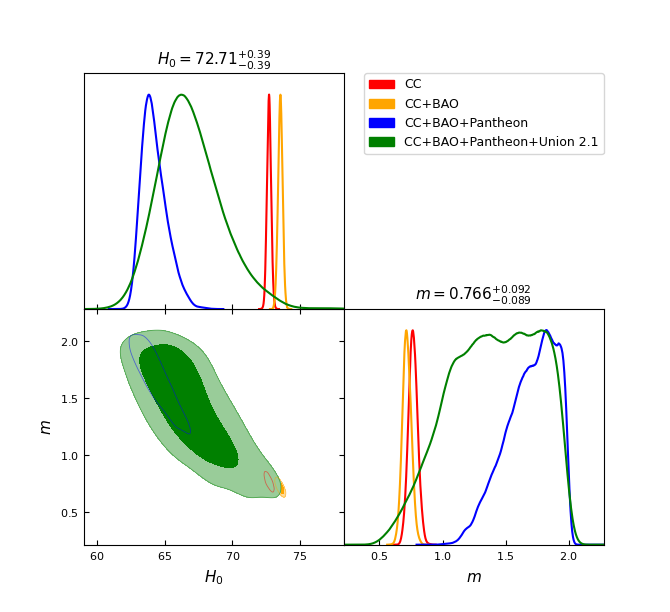}
\caption{\label{Q^(-2)T^2(4_legended_plot)}Posterior distribution of model parameter $m$ using CC, BAO, SN Ia data. The deeper shade show 68\% credible level (C.L.) and the lighter shade show 95\% credible level (C.L.)} 
\end{figure}
\begin{table*}[h!]
\caption{\label{Table_5}Best fit parameter values for Model V from MCMC}
\vspace{2mm}
\begin{tabular}{m{7cm} m{4.8cm} m{1.7cm}}
\hline
 Datasets & $H_{0}(km/s/Mpc)$ &  $m$  \\ 
\hline 
\vspace{1mm}
Priors & (60,80) & (0,2)\\
\vspace{1mm}
 CC & $66.7_{-4.8}^{+6.5}$ & $1.45_{-0.80}^{+0.59}$  \\
\vspace{1mm}
 CC+BAO & $64.3_{-1.7}^{+2.8}$ & $1.70_{-0.51}^{+0.33}$  \\
 \vspace{1mm}
 CC+BAO+Pantheon & $73.54_{-0.47}^{+0.44}$ & $0.720_{-0.089}^{+0.1}$ \\
 \vspace{1mm}
CC+BAO+Pantheon+Union & $72.71_{-0.39}^{+0.39}$ & $0.766_{-0.089}^{+0.092}$ \\
\hline
\end{tabular}
\end{table*}

\begin{figure}[htbp]
    \centering
    \begin{minipage}[b]{0.46\textwidth}
        \centering
        \includegraphics[width=\textwidth]{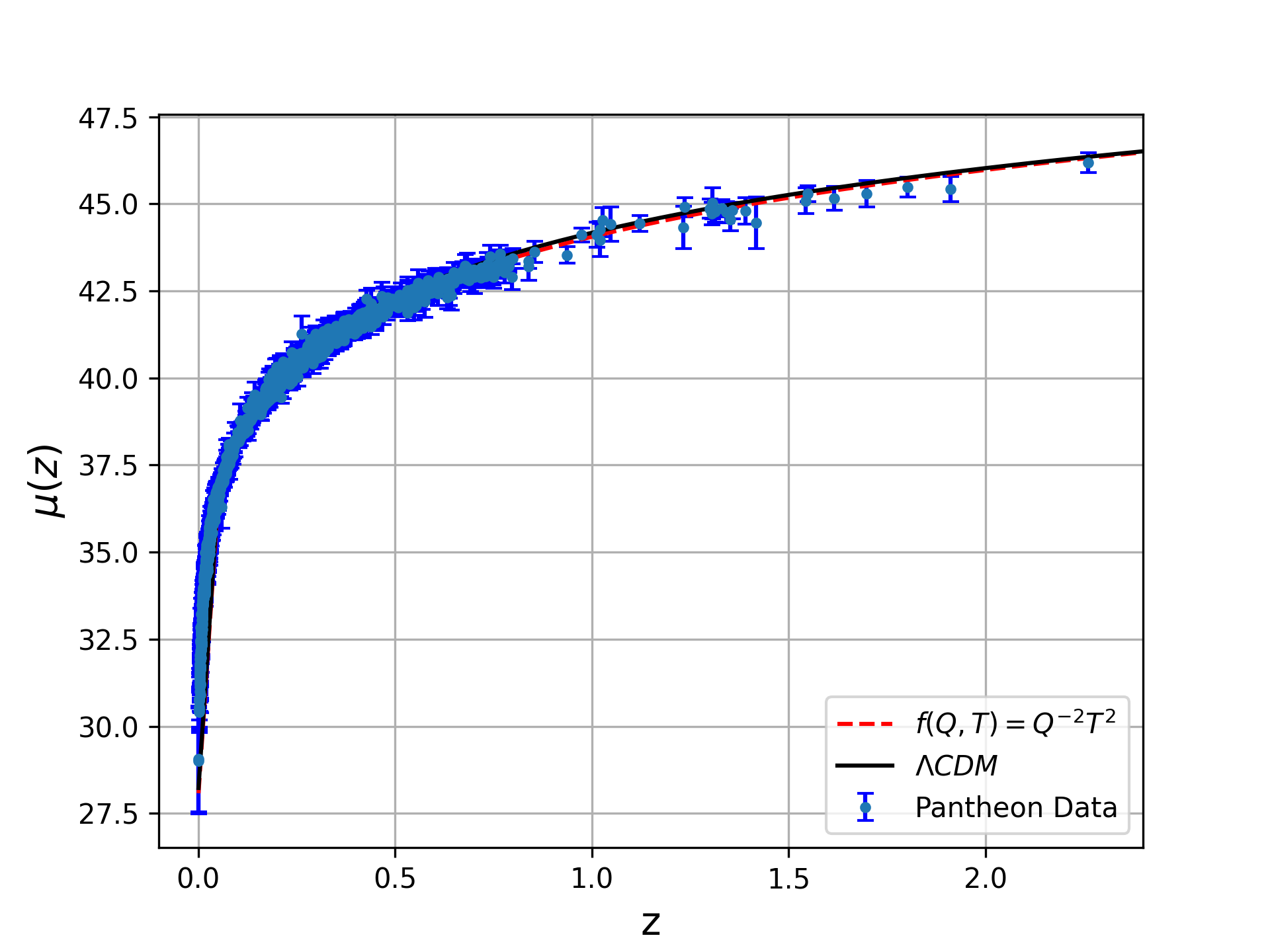}
        \caption{Plot of distance modulus $\mu(z)$ vs. redshift $z$ for Model V shown in red dotted line and $\Lambda CDM$ in black solid line. Our model shows an excellent fit with the 1701 points of Pantheon dataset\cite{brout2022pantheon+}.}
        \label{distance modulus(Q^(-2)T^2)}
    \end{minipage}
    \hspace{0.05\textwidth}
    \begin{minipage}[b]{0.46\textwidth}
        \centering
        \includegraphics[width=\textwidth]{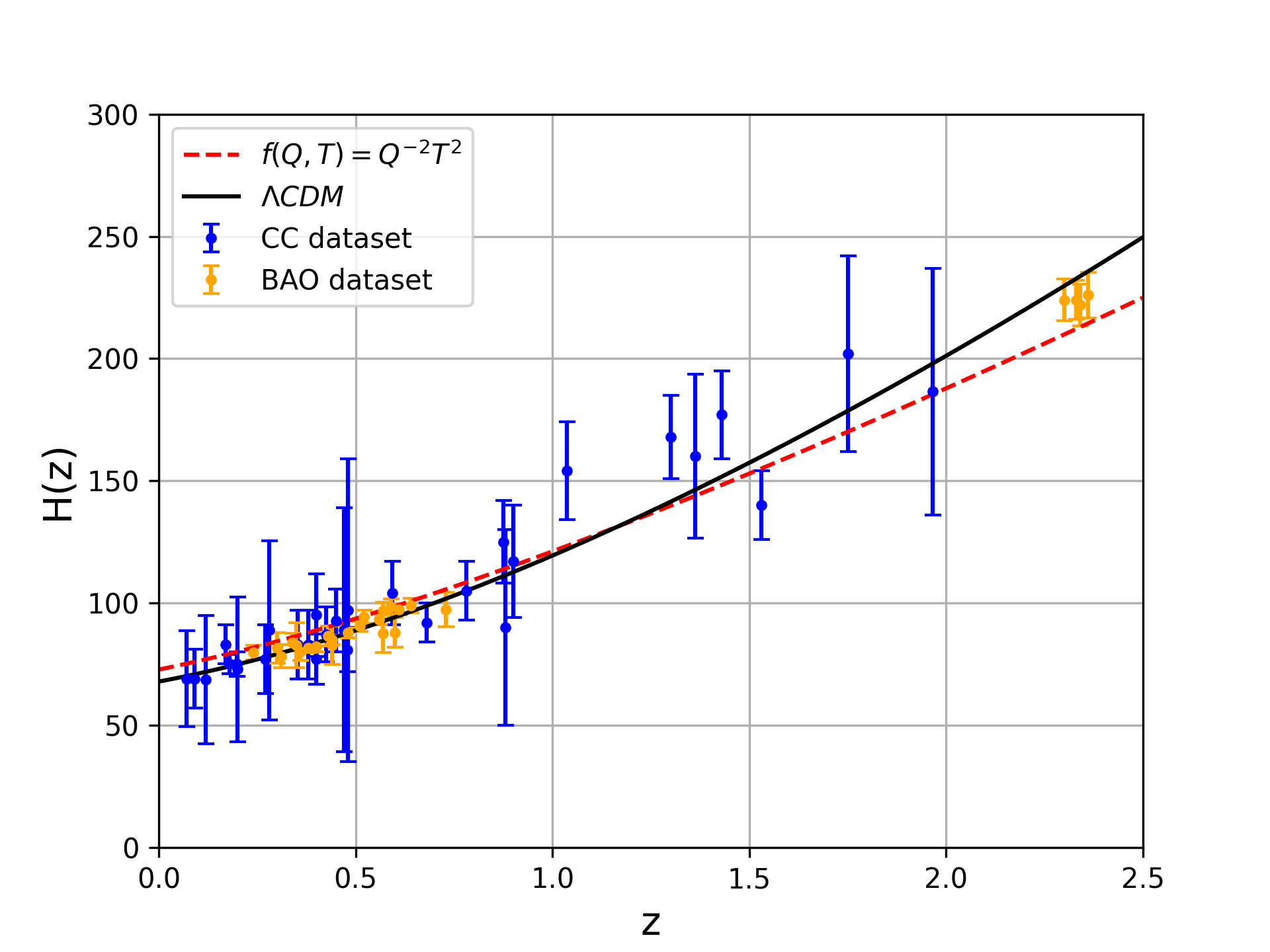}
        \caption{Comparison of model V for best-fit parameter value, which is shown in red dotted line and 57 Hubble data points\cite{yu2018hubble} along with $\Lambda CDM$ model which is shown in black solid line.}
        \label{H(z)(Q^(-2)T^(2))}
    \end{minipage}
\end{figure}

By solving this model numerically we have shown the best fit values for the parameters of this model in Fig.\ref{Q^(-2)T^2(4_legended_plot)} contour plot. The best fit value of the model parameters are shown in Table \ref{Table_5} for different datasets and the joint analysis. Our present non-minimally coupled model gives a value of $H_0$ as $73.54_{-0.47}^{+0.44}$ km/s/Mpc (from joint analysis of CC+BAO+Pantheon), thus successfully reducing the Hubble tension.  In Fig.\ref{distance modulus(Q^(-2)T^2)} we have plotted our model along with the $\Lambda CDM$ model with the best fit parameter values obtained from MCMC. The plot also displays the Pantheon+ SNe Ia findings, 1701 data points, with their errors, allowing for a direct comparison of the two models.

\emph{Other Cosmological Behaviours:}

Let us now proceed to the numerical investigation of this model. Similarly to the previous Model IV, we set the same constants for the $\Lambda CDM$ model.\\

\begin{figure}[htbp]
    \centering
    \begin{minipage}[b]{0.45\textwidth}
        \centering
        \includegraphics[width=\textwidth]{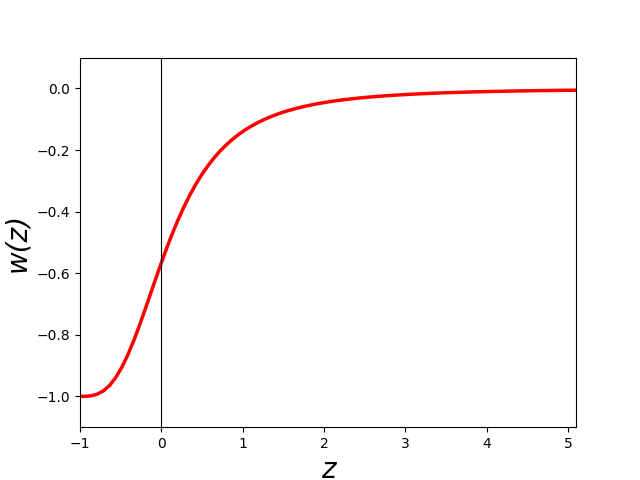}
        \caption{Variation of EoS parameter for the best fit value of $m=0.766$ from the MCMC analysis of four datasets}
        \label{w(z)(Q^(-2)T^2)}
    \end{minipage}
    \hspace{0.05\textwidth}
    \begin{minipage}[b]{0.45\textwidth}
        \centering
        \includegraphics[width=\textwidth]{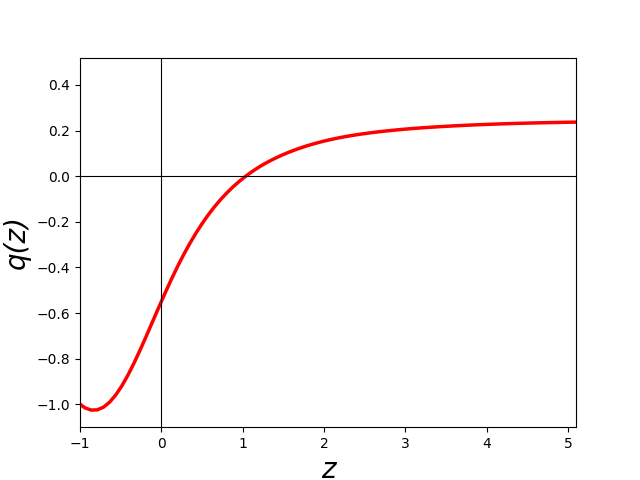}
        \caption{Variation of deceleration parameter $q(z)$ for the best fit value of $m=0.766$ from the joint analysis of four datasets}
        \label{q(z)(Q^(-2)T2)}
    \end{minipage}
\end{figure}
In Fig.\ref{H(z)(Q^(-2)T^(2))} we present the evolution of Hubble function by solving Eq.(\ref{70}) with respect to redshift $z$ by using the best fit parameter value shown in Fig.\ref{Q^(-2)T^2(4_legended_plot)}  along with the $\Lambda CDM$ model and the observed data on $H(z)$, accompanied by error bars. The agreement between the model prediction and observed data is similar to that of Model IV. It agrees fairly well with the data with slight disagreement in the high redshift region.\\
Now, we have solved this model numerically and shown the behavior of EoS parameter in Fig.\ref{w(z)(Q^(-2)T^2)} for the best fit parameter value $m$. The value of EoS parameter at present day($z=0$) is $\omega_{0}=-0.56_{-0.03}^{+0.02}$ which indicates an accelerating phase of the universe.
The behavior of deceleration parameter is shown in Fig.\ref{q(z)(Q^(-2)T2)} for the best fit parameter values of $\alpha$,$\beta$ and $m$. We found that there is a well behaved transition from deceleration to acceleration phase and the present day value of $q$ is $q_{0}=-0.53_{-0.04}^{+0.03}$ which is negative at present time indicating the accelerating phase of the universe.

\section{Gravitational wave analysis}
\label{sec5}
In the previous section, we introduced a modified theory of gravity in which the propagation of GW signals, along with the corresponding luminosity distance, $d_{L}^{gw}$, differs from predictions based on standard Einstein theory. In this section, we will explore the possibility that this modified GW dynamics might be consistent with current cosmological observations. Assuming that the Universe is homogeneous and isotropic on large scales, the FLRW metric, as expressed in Eq.(\ref{9}), is used to describe its geometry. Within this metric, and in the framework of $f(Q,T)$ gravity theory, the GW luminosity distance $d_{L}^{gw}$ can be formulated as shown in Eq.(\ref{29}). Our analysis is based on the data from the Gravitational-Wave-Transient-Catalog (GWTC), which includes compact binary coalescences observed by the LIGO-VIRGO and KAGRA collaborations\cite{abbott2019gwtc,abbott2021gwtc,abbott2023gwtc,abbott2024gwtc} during their various observing runs.

\subsection{Model I : $f(Q,T) =Q+\beta T$}
The modified GW luminosity distance Eq.(\ref{29}), for Model I, takes the form,
\begin{equation}\label{71}
    d_{L}^{gw}=c(1+z)\int_{0}^{z}\frac{dz'}{H(z')}
\end{equation}
As we can see, there is no deviation of $d_{L}^{gw}$ from $d_{L}^{em}$ , which was expected in the presence of linear dependence on $Q$. For an explicit evaluation of the GW luminosity distance, we need to calculate $H(z)$, which has been determined in Eq.(\ref{45}) by solving the Friedmann equations (\ref{10}) and (\ref{11}). In addition, and for a numerical estimate of the differences with standard cosmological model($\Lambda CDM$), we have used the result of the joint analysis of four datasets namely CC, BAO, Pantheon and Union 2.1  of MCMC for this model.\\

\begin{figure}[h!]
    \centering
    \includegraphics[scale=0.8]{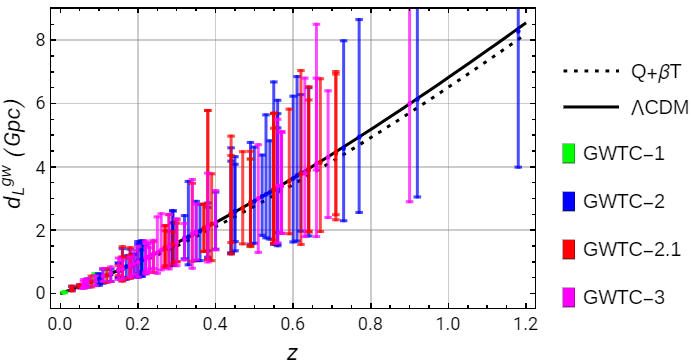}
    \caption{The plot illustrates the comparison between the GW luminosity distance for the best-fit parameter value and the observed data. The solid black line corresponds to the $\Lambda$CDM model, while the black dotted line represents our $f(Q,T)$ Model I. The colored error bars indicate data from various observing runs conducted by the Advanced LIGO and VIRGO observatories\cite{abbott2019gwtc,abbott2021gwtc,abbott2023gwtc,abbott2024gwtc}.}
    \label{dL(gw)(n=0)}
\end{figure}

In Fig.\ref{dL(gw)(n=0)}, we have shown the GW luminosity distance along with $\Lambda CDM$ model and the observational data. The figure shows a very good match of our model with $\Lambda CDM$ model and LIGO-VIRGO data. 

\subsection{Model II : $f(Q,T) =Q^{2}+\beta T$}
For the second example, the modified luminosity distance takes the form,
\begin{equation}\label{72}
    d_{L}^{gw}=c(1+z)\left(\frac{H(0)}{H(z)}\right)\int_{0}^{z}\frac{dz'}{H(z')}
\end{equation}
where $H(z)$ is given by the Eq.(\ref{52}) and $H(0)$ is the present day value of Hubble parameter given by this model.
Again by using the similar technique as Model I, we have used the result of MCMC of joint analysis of the four datasets for this model.
\begin{figure}[h!]
    \centering
    \includegraphics[scale=0.8]{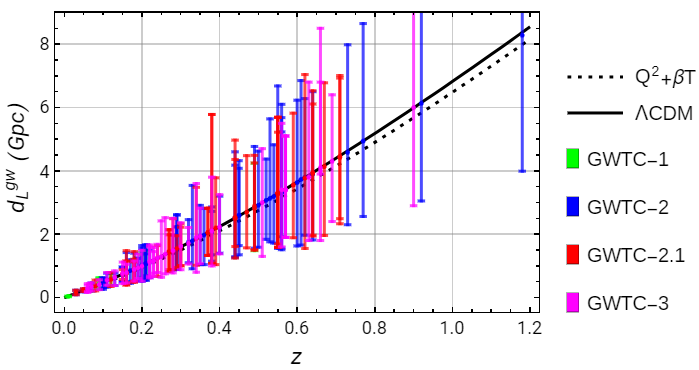}
    \caption{The plot displays the comparison between the GW luminosity distance and observational data. The solid black line corresponds to the $\Lambda$CDM cosmological model and the black dotted line illustrates the predictions of our $f(Q,T)$ Model II. The colored error bars indicate observational data from various observing runs of the Advanced LIGO and VIRGO collaborations\cite{abbott2019gwtc,abbott2021gwtc,abbott2023gwtc,abbott2024gwtc}.}
    \label{dL(gw)(n=1)}
\end{figure}
In Fig.\ref{dL(gw)(n=1)}, we have shown the modified GW luminosity distance along with $\Lambda CDM$ model and the observational data. The figure shows an excellent match of our model with LIGO-VIRGO data.

\subsection{Model III : $f(Q,T) =Q^{3} +\beta T$}
For the third example, we consider the form $f(Q,T) =Q^{3}+\beta T$, where $\beta$ is a constant. For the present case, the modified luminosity distance takes the form,
\begin{equation}\label{73}
    d_{L}^{gw}=c(1+z)\left(\frac{H(0)}{H(z)}\right)^{2}\int_{0}^{z}\frac{dz'}{H(z')}
\end{equation}
where $H(z)$ is given by the Eq.(\ref{57}) and $H(0)$ is the present day value of Hubble parameter given by this model.
Using the similar technique, we applied the results of the MCMC joint analysis of the four datasets for this model.

\begin{figure}[h!]
    \centering
    \includegraphics[scale=0.8]{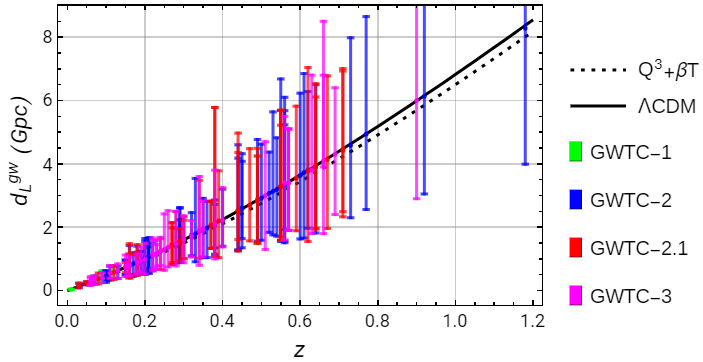}
    \caption{The plot between GW luminosity distance and redshift. The solid black line represents the $\Lambda CDM$ scenario while the black dotted line represents our $f(Q,T)$ model III, provides us a direct comparison between two models. The colored error bars represents different observing runs of Advanced LIGO and VIRGO observatories\cite{abbott2019gwtc,abbott2021gwtc,abbott2023gwtc,abbott2024gwtc}.}
    \label{dL(gw)(n=2)}
\end{figure}
In Fig.\ref{dL(gw)(n=2)}, we have shown the modified GW luminosity distance along with $\Lambda CDM$ model and the observational data. The figure shows an excellent match of our model with $\Lambda CDM$ model.

\subsection{Model IV : $f(Q,T) =-\alpha Q -\beta T^{2}$}
For the fourth example, we consider the form $f(Q,T) =-\alpha Q -\beta T^{2}$, where $\alpha$, $\beta$ are constants. For the present case, the modified luminosity distance takes the form,
\begin{equation}\label{74}
    d_{L}^{gw}=c(1+z)\int_{0}^{z}\frac{dz'}{H(z')}
\end{equation}
where $H(z)$ is given by the solution of Eq.(\ref{64}) and $H(0)$ is the present day value of Hubble parameter given by this model.
Similar to the approach used in Model I, we applied the results from the MCMC joint analysis of the four datasets to this model.

\begin{figure}[h!]
    \centering
    \includegraphics[scale=0.8]{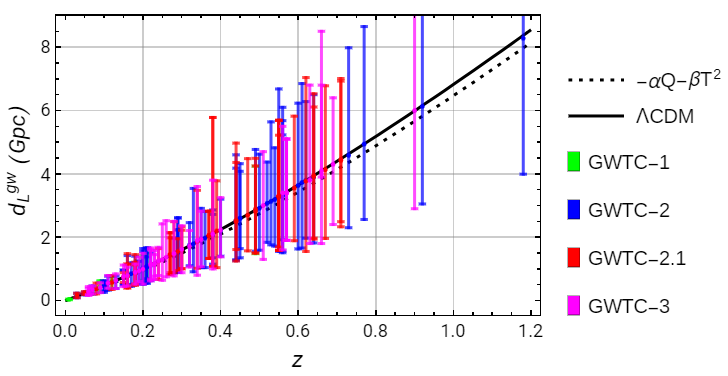}
    \caption{The plot between GW luminosity distance and redshift. The solid black line represents the $\Lambda CDM$ scenario while the black dotted line represents our $f(Q,T)$ model IV. The colored error bars represents data of different observing runs of Advanced LIGO and VIRGO observatories\cite{abbott2019gwtc,abbott2021gwtc,abbott2023gwtc,abbott2024gwtc}.}
    \label{dL(gw)(-aQ-bT2)}
\end{figure}
In Fig.\ref{dL(gw)(-aQ-bT2)}, we have plotted the modified GW luminosity distance along with $\Lambda CDM$ model and the observational data. The figure shows an excellent match of our model with LIGO-VIRGO data.

\subsection{Model V : $f(Q,T) =Q^{-2}T^{2}$}
For the fifth example, we consider the form $f(Q,T) =Q^{-2}T^{2}$. For the present case, the modified luminosity distance takes the form,
\begin{equation}\label{75}
    d_{L}^{gw}=c(1+z)\left(\frac{H(0)}{H(z)}\right)^{2}\int_{0}^{z}\frac{dz'}{H(z')}
\end{equation}
where $H(z)$ is given by the solution of Eq.(\ref{70}) and $H(0)$ is the present day value of Hubble parameter given by this model.
Again by following the similar technique of previous models, we have used the result of MCMC of joint analysis of the four datasets for this model.

\begin{figure}[h!]
    \centering
    \includegraphics[scale=0.8]{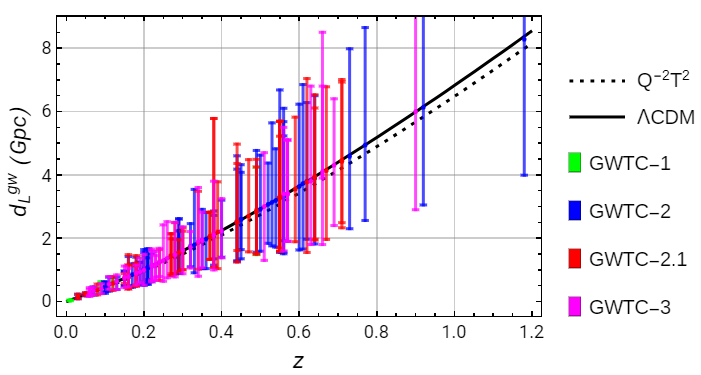}
    \caption{The plot between GW luminosity distance in gigapersec unit and redshift. The solid black line represents the $\Lambda CDM$ scenario while the dotted line represents our model V. The observational data is represented by different colored points and with their error bars\cite{abbott2019gwtc,abbott2021gwtc,abbott2023gwtc,abbott2024gwtc}.}
    \label{dL(gw)(Q^(-2)T^2)}
\end{figure}
\begin{figure}[h!]
    \centering
    \includegraphics[scale=0.5]{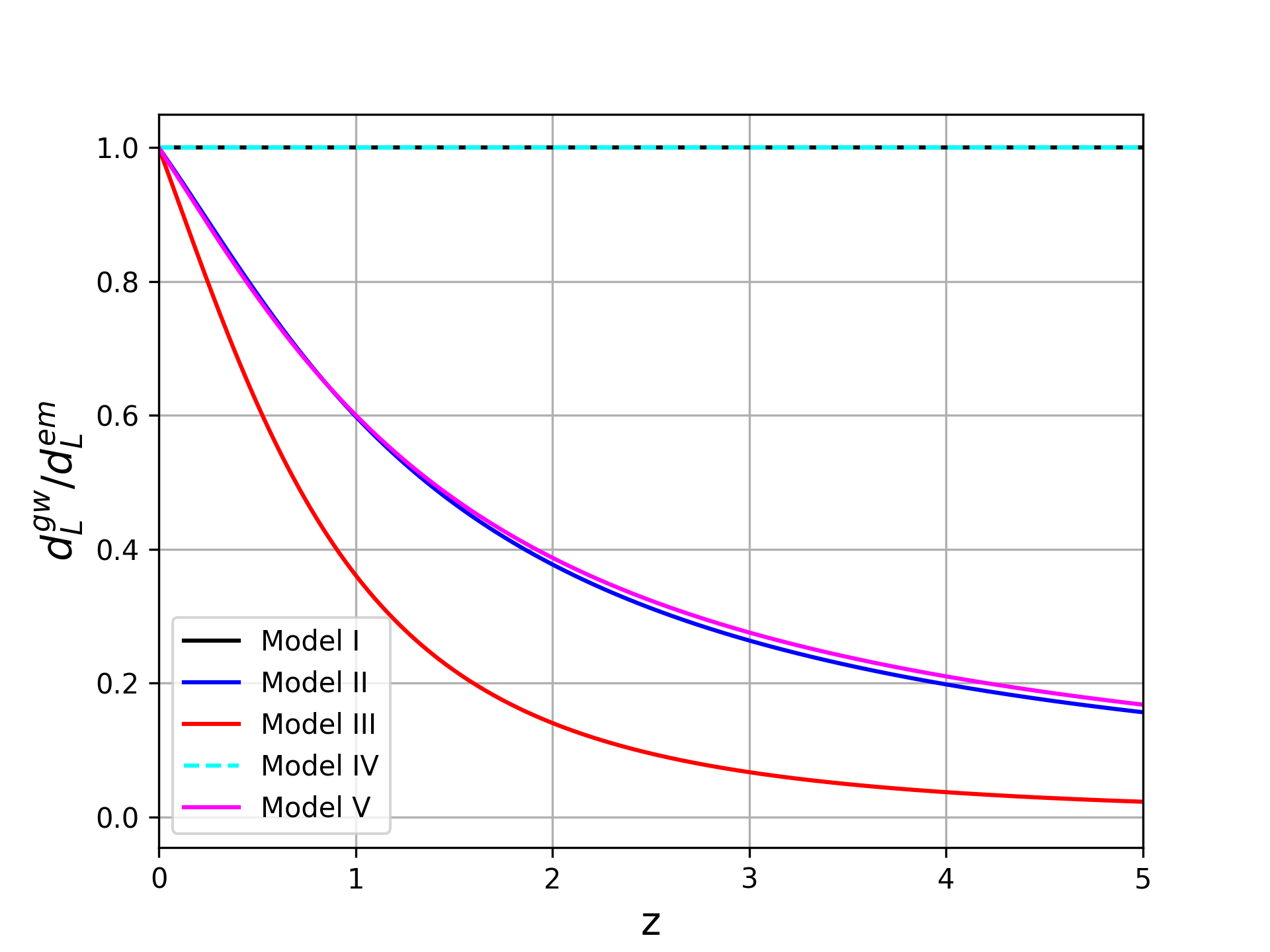}
    \caption{The plot shows the evolution of the ratio $d_{L}^{gw}/d_{L}^{em}$ as a function of redshift, with each colored line representing deviations from GR for various $f(Q,T)$ gravity models.}
    \label{dL(gw)_em}
\end{figure}

In Fig.\ref{dL(gw)(Q^(-2)T^2)}, we have plotted the modified GW luminosity distance along with $\Lambda CDM$ model and the observational data. The figure shows a very good match of our model with $\Lambda CDM$ model.\\
In Fig.\ref{dL(gw)_em} we have plotted the ratio $d_{L}^{gw}/d_{L}^{em}$ as a function of $z$ for the best fit parameter values for our assumed models in order to show the deviation of $f(Q,T)$ gravity models from GR. As we can see, the deviations increase with redshift. The ratio $d_L^{gw}/d_L^{em}$ turn out to be less than $1$ for the models II, III and V (being maximum for model III) and matches with GR for models I and IV, which is also evident from the corresponding $d_L^{gw}$ expressions. This means that for models II, III and V, received GW signals turn out to be stronger with respect to the value predicted by the standard Einstein theory and thus will be more easier to be detected, given the same distance and detection sensitivity. 

\begin{figure}[h!]
    \centering
    \includegraphics[scale=0.8]{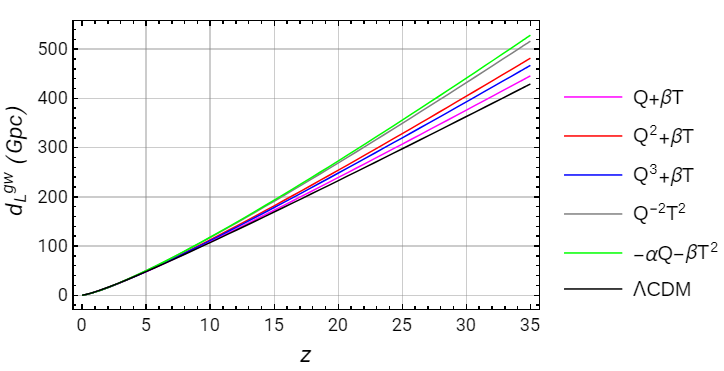}
    \caption{Comparison of GW luminosity distance for all models along with $\Lambda CDM$. The deviation from the $\Lambda CDM$ can be seen at very high redshift.}
    \label{dL(gw)All_model}
\end{figure}

In Fig.\ref{dL(gw)All_model} we have shown the variation of $d_L^{gw}$ with redshift for the best fit parameter values for all the models along with $\Lambda$CDM for comparison. We can see that though the $d_L^{gw}$ for the models show similar behaviour for small redshifts, their deviations become more prominent as we go to larger $z$, being maximum for Model IV. 

\section{Conclusions}
\label{sec6}

\begin{table*}
\caption{\label{Table_6}Best fit parameter values for our models from MCMC}
\vspace{2mm}
\begin{tabular}{m{2.5cm} m{2cm} m{1.5cm} m{2cm} m{2cm} m{2cm} m{1.5cm}}
\hline
 Models & $H_{0}(\frac{km/s}{Mpc})$ &  $\alpha$ & $\beta$ & $m$ & $w_0$ & $q_0$ \\ 
\hline 
\vspace{2mm}
$Q+\beta T$ & $73.06_{-0.38}^{+0.38}$ & - & $2.6_{-3.4}^{+1.5}$ & $0.387_{-0.088}^{+0.057}$ & $-0.72_{-0.05}^{+0.03}$ & $-0.61_{-0.03}^{+0.03}$\\
\vspace{2mm}
 $Q^{2}+\beta T$ & $72.90_{-0.40}^{+0.41}$ & - & $-11.64_{-0.40}^{+0.71}$ & $0.112_{-0.054}^{+0.1}$ & $-0.89_{-0.05}^{+0.07}$ & $-0.58_{-0.03}^{+0.02}$\\
\vspace{2mm}
 $Q^{3}+\beta T$ & $72.98_{-0.42}^{+0.41}$ & - & $-14.05_{-0.34}^{+0.39}$ & $-0.305_{-0.045}^{+0.058}$ & $-1.44_{-0.1}^{+0.11}$ & $-0.59_{-0.03}^{+0.01}$\\
 \vspace{2mm}
 $-\alpha Q-\beta T^{2}$ & $72.76_{-0.41}^{+0.40}$ & $3.4_{-4.5}^{+5.2}$ & $3.4_{-4.6}^{+5.2}$ & $0.458_{-0.091}^{+0.062}$ & $-0.68_{-0.05}^{+0.02}$ & $-0.54_{-0.05}^{+0.03}$ \\
 \vspace{2mm}
$Q^{-2}T^{2}$ & $72.71_{-0.39}^{+0.39}$ & - & - & $0.766_{-0.089}^{+0.092}$ & $-0.56_{-0.03}^{+0.02}$ & $-0.53_{-0.04}^{+0.03}$ \\
\hline
\end{tabular}
\end{table*}

As new theories of gravity emerge, it is essential to rigorously test their viability in explaining the universe's dark sector. The $f(Q,T)$ gravity theory, which combines the non-metricity function $Q$ with the trace of the energy-momentum tensor $T$, presents a promising approach. In this study, we evaluate several viable models of $f(Q,T)$ gravity against both cosmological data and gravitational wave observations to assess their potential.\\
To begin with, we considered minimally coupled models having both linear and non linear dependence on $Q, \ T$, having the general form $f(Q,T)=Q^n+\beta T$. We study the cases corresponding to $n=1,\ 2,\ 3$. Model I has the functional form $f(Q,T)=Q+\beta T$, where $\beta$ is a free parameter.
We used a widely accepted parametric form of the equation of state parameter as a function of redshift $z$ to solve the field equations for the Hubble parameter $H$. This parameter exhibits a negative value during the recent epoch of acceleration. At high redshift $z$, $\omega$ approaches zero for positive values of the model parameter $m$, while its value at $z=0$ depends on this parameter. To test our model, we utilized three datasets: the Hubble dataset comprising 57 data points (CC+BAO), the Pantheon dataset with 1701 data points, and the Union 2.1 dataset containing 580 data points. Then we employed MCMC method to put stringent constraints on all the model parameters by using all four datasets (CC+BAO+Pantheon+Union 2.1) individually and jointly. The best fit values of our model parameters from our study of the joint analysis (CC+BAO+Pantheon+Union 2.1) are as follows: $\beta=2.6_{-3.4}^{+1.5}$\;,$m=0.387_{-0.088}^{+0.057}$. For these best fit values, our model gives the present day value of Hubble parameter as $H_{0}=73.06_{-0.38}^{+0.38}$km/s/Mpc. While the joint analysis (CC+BAO+Pantheon) gives a value of $H_0$ as $H_{0}=73.97_{-0.45}^{+0.44}$ km/s/Mpc which is consistent with the latest direct measurement of Hubble parameter. For these constant values we have tested our model with GW luminosity distance data points obtained from different observational runs of LIGO-VIRGO and KAGRA collaboration along with $\Lambda CDM$ model. Fig.\ref{dL(gw)(n=0)} shows quite good match with observational data points. We then examined the behavior of EoS parameter and found the present day value is $\omega_{0}=-0.72_{-0.05}^{+0.03}$ which indicates an accelerating phase. Then we also studied the deceleration parameter and this model predicts the present day value as $q_{0}=-0.61_{-0.03}^{+0.03}$ which is negative at present time indicating the accelerating phase of the universe.\\
For our second example, we considered the form $f(Q,T)=Q^{2}+\beta T$, where again $\beta$ is a free parameter. Similar to the previous model, we used the same parametric form of equation of state parameter to solve the field equations for $H$. Then we employed MCMC method to put stringent constraints on the model parameters. The parameters from our study are as follows: $\beta=-11.64_{-0.40}^{+0.71}$\;,$m=0.112_{-0.054}^{+0.10}$ obtained from joint analysis of all the 4 datasets. For these best fit values, our model gives the present day value of Hubble parameter as $H_{0}=72.90_{-0.40}^{+0.41}$km/s/Mpcs. While from the joint analysis of (CC+BAO+Pantheon), $H_{0}=73.79_{-0.47}^{+0.45}$ km/s/Mpc which is consistent with the observational data. For these constant values we have tested our model with GW luminosity distance data points along with $\Lambda CDM$ model. Fig.\ref{dL(gw)(n=1)} shows quite good match with observational data points. We then examined the behavior of EoS parameter and found the present day value is $\omega_{0}=-0.89_{-0.05}^{+0.07}$ which indicates an accelerating phase. Then we also studied the deceleration parameter and the present day value is $q_{0}=-0.58_{-0.03}^{+0.02}$.\\
Similarly, for our third example of the form $f(Q,T)=Q^{3}+\beta T$, we have got the parameter values as $\beta=-14.05_{-0.34}^{+0.39}$\;,$m=-0.305_{-0.045}^{+0.058}$. For these best values, our model gives the present day value of Hubble parameter as $H_{0}=72.98_{-0.42}^{+0.41}$km/s/Mpc which is again an improvement over the value of $H_0$ as obtained from $\Lambda$CDM model. For these constant values we have tested our model with GW luminosity distance data points along with $\Lambda CDM$ model. Fig.\ref{dL(gw)(n=2)} shows a good agreement with observational data points. We then examined the behavior of EoS parameter and found the present day value is $\omega_{0}=-1.44_{-0.1}^{+0.11}$ which indicates an accelerating phase. However, we found that a negative value of $m$ results in a sudden transition of EoS from negative to positive value in and around $z=1$, beyond which it tends toward $\omega=0$. Therefore, the EoS doesn't give the correct behaviour and hence is not acceptable, thereby disfavouring this model over the others. Then we also studied the deceleration parameter and the present day value is $q_{0}=-0.59_{-0.03}^{+0.01}$ which is negative at present time indicating the accelerating phase of the universe.\\
Finally we considered a fourth functional form for minimally coupled model where we introduced the non-linearity in $T$. We considered the form $f(Q,T)=-\alpha Q-\beta T^{2}$, where $\alpha$ and $\beta$ are free, positive parameters. Then we employed MCMC method to put constraint on the model parameters and we found $\alpha=3.4_{-4.5}^{+5.2}$\;,$\beta=3.4_{-4.6}^{+5.2}$\;,$m=0.458_{-0.091}^{+0.062}$. For these best fit parameters, our model gives the Hubble parameter value as $H_{0}=72.76_{-0.41}^{+0.40}$ which is again shows the agreement with observational data. For these constant values we have tested our model with GW luminosity distance data points along with $\Lambda CDM$ model. Fig.\ref{dL(gw)(-aQ-bT2)} shows a good agreement with observational data points. We then examined the behavior of EoS parameter and found the present day value is $\omega_{0}=-0.68_{-0.05}^{+0.02}$ which indicates an accelerating phase. Then we also studied the deceleration parameter and the present day value is $q_{0}=-0.54_{-0.05}^{+0.03}$ which is negative at present time indicating the accelerating phase of the universe.\\
For our last model, we considered a particular example of non-minimally coupled model of the form $f(Q,T)=Q^{-2}T^{2}$. By employing MCMC method we found the constant values as: $H_{0}=72.71_{-0.39}^{+0.39}$\;,$m=0.766_{-0.089}^{+0.092}$. For these constant values we have tested our model with GW luminosity distance data points along with $\Lambda CDM$ model. Fig.\ref{dL(gw)(Q^(-2)T^2)} shows a good match with observational data points. We then examined the behavior of EoS parameter and found the present day value is $\omega_{0}=-0.56_{-0.03}^{+0.02}$ which indicates an accelerating phase. Then we also studied the deceleration parameter and the present day value is $q_{0}=-0.53_{-0.04}^{+0.03}$ which is negative at present time indicating the accelerating phase of the universe.

We further showed that the above models fit quite well with the latest gravitational wave LIGO-VIRGO data based on the study of the modified gravitational wave luminosity distance as discussed in Section \ref{sec5}. In Fig. \ref{dL(gw)All_model} we also found that though the $d_L^{gw}$ of these models show similar behaviour and follow closely $\Lambda$CDM model for low redshifts, they show significant deviations among each other and from $\Lambda$CDM as we go to high redshifts. Thus, future gravitational wave data for higher redshifts may provide grounds to further falsify these models.

From the above results, we can clearly conclude that the Hubble parameter acquires increased value in and around the present time in the presence of  $f(Q,T)$ like background dynamics.The additional terms coming from this model lead to the alleviation of the existing Hubble tension without the need for any cosmological constant/dark energy. We show that there are a wide number of combinations of $Q,T$ that can lead to a well behaved cosmological model that satisfies a wide range of cosmological and gravitational wave data. 

In Table \eqref{Table_6}, we have quoted the best fit values of our model parameters $(\alpha,\ \beta,\ m)$ and the present values of $(H_0,\ w_0,\ q_0)$ that we have obtained for the above mentioned models for comparison.

In a following work we plan to further test these models without the assumption of any particular parametrization of the EoS $\omega$. We further plan to test these models in the context of third generation gravitational wave detectors like Einstein Telescope, cosmic explorers through the simulation of a series of NS-NS and NS-BH merger events. We also plan to study structure formation for these models and examine the $\sigma_8$ tension.

\begin{acknowledgments} 
SB and AP would like to thank the department of physics, IIT(ISM) Dhanbad for their support during the completion of this project.

\end{acknowledgments}

\bibliographystyle{JHEP}
\bibliography{mybib}

\end{document}

%% file: newcommands.tex
\let\oldsqrt\sqrt
\def\sqrt{\mathpalette\DHLhksqrt}
\def\DHLhksqrt#1#2{%
\setbox0=\hbox{$#1\oldsqrt{#2\,}$}\dimen0=\ht0
\advance\dimen0-0.2\ht0
\setbox2=\hbox{\vrule height\ht0 depth -\dimen0}%
{\box0\lower0.4pt\box2}}

\newcommand{\beq}{\begin{equation}}
\newcommand{\eeq}{\end{equation}}
\newcommand{\bea}{\begin{equation}\begin{aligned}}
\newcommand{\eea}{\end{aligned}\end{equation}}

\newlength{\wsingfig}
\newlength{\wdblefig}
\newlength{\wquadfig}
\newlength{\wtriplefig}
